\documentclass[acmsmall]{acmart}

\AtBeginDocument{%
  }

\setcopyright{cc}
\setcctype{by}
\acmJournal{PACMHCI}
\acmYear{2025} \acmVolume{9} \acmNumber{2} \acmArticle{CSCW132} \acmMonth{4} \acmPrice{}\acmDOI{10.1145/3711030}

\received{January 2024}
\received[revised]{July 2024}
\received[accepted]{October 2024}
\usepackage{longtable}
\usepackage{array}
\usepackage{wrapfig}
\usepackage{multirow}
\usepackage{subcaption}

\begin{document}

% \title[MeetMap]{MeetMap: Real-Time Capture of Verbal Communication in Meetings Through Human-AI Interactive Diagrams }
\title[MeetMap: Real-Time Collaborative Dialogue Mapping with LLMs in Online Meetings]{MeetMap: Real-Time Collaborative Dialogue Mapping with LLMs in Online Meetings}

\author{Xinyue Chen}
\email{xinyuech@umich.edu}
\orcid{0000-0002-0774-223X}
\affiliation{%
  \institution{University of Michigan}
  \state{Michigan}
  \country{USA}
}

\author{Nathan Yap}
\email{nyap@umich.edu}
\orcid{0009-0003-7760-1496}
\affiliation{%
  \institution{University of Michigan}
  \state{Michigan}
  \country{USA}
}

\author{Xinyi Lu}
\email{lwlxy@umich.edu}
\orcid{0000-0003-2610-6084}
\affiliation{%
  \institution{University of Michigan}
  \state{Michigan}
  \country{USA}
}

\author{Aylin Gunal}
\email{gunala@umich.edu}
\orcid{0009-0005-4028-9405}
\affiliation{%
  \institution{University of Michigan}
  \state{Michigan}
  \country{USA}
}

\author{Xu Wang}
\email{xwanghci@umich.edu}
\orcid{0000-0001-5551-0815}
\affiliation{%
  \institution{University of Michigan}
  \state{Michigan}
  \country{USA}
}

% \affiliation{%

\newcommand{\xw}[1]
{{\fontfamily{cmss}\selectfont \bfseries \color{blue} XW: #1 }}

\newcommand{\xy}[1]
{{\fontfamily{cmss}\selectfont \bfseries \color{blue}  #1 }}

\definecolor{HumanMap}{HTML}{abdbe3}
\definecolor{AIMap}{HTML}{e3b3ab}
\definecolor{Baseline}{HTML}{FFFFFF}
\setlength{\fboxsep}{1pt}

\newcommand{\HumanMap}{\textcolor{black}{\fcolorbox{HumanMap}{HumanMap}{Human-Map}} }
\newcommand{\AIMap}{\textcolor{black}{\fcolorbox{AIMap}{AIMap}{AI-Map}} }
\newcommand{\Baseline}{\textcolor{black}{\fcolorbox{Baseline}{Baseline}{baseline}} }

\newcommand{\revise}[1]
{{\color{orange}  #1 }}

\renewcommand{\shortauthors}{Xinyue Chen, Nathan Yap, Xinyi Lu, Aylin Gunal, and Xu Wang}
\begin{abstract}
Video meeting platforms display conversations linearly through transcripts or summaries. However, ideas during a meeting do not emerge linearly. We leverage LLMs to create dialogue maps in real time to help people visually structure and connect ideas.
Balancing the need to reduce the cognitive load on users during the conversation while giving them sufficient control when using AI, we explore two system variants that encompass different levels of AI assistance. In Human-Map, AI generates summaries of conversations as nodes, and users create dialogue maps with the nodes. In AI-Map, AI produces dialogue maps where users can make edits. 
We ran a within-subject experiment with ten pairs of users, comparing the two MeetMap variants and a baseline. Users preferred MeetMap over traditional methods for taking notes, which aligned better with their mental models of conversations. Users liked the ease of use for AI-Map due to the low effort demands and appreciated the hands-on opportunity in Human-Map for sense-making.  %This work informs the future design of AI-assisted tools for real-time cognitive scaffolding in meetings by emphasizing the necessity to balance AI assistance with synchronicity and user agency to enhance collaborative sense-making.
% This work guides the design of AI tools for real-time cognitive support in meetings, highlighting the need to balance AI assistance with synchronicity and user agency.
\end{abstract}







\begin{teaserfigure}
    \centering
    \includegraphics[width=\textwidth]{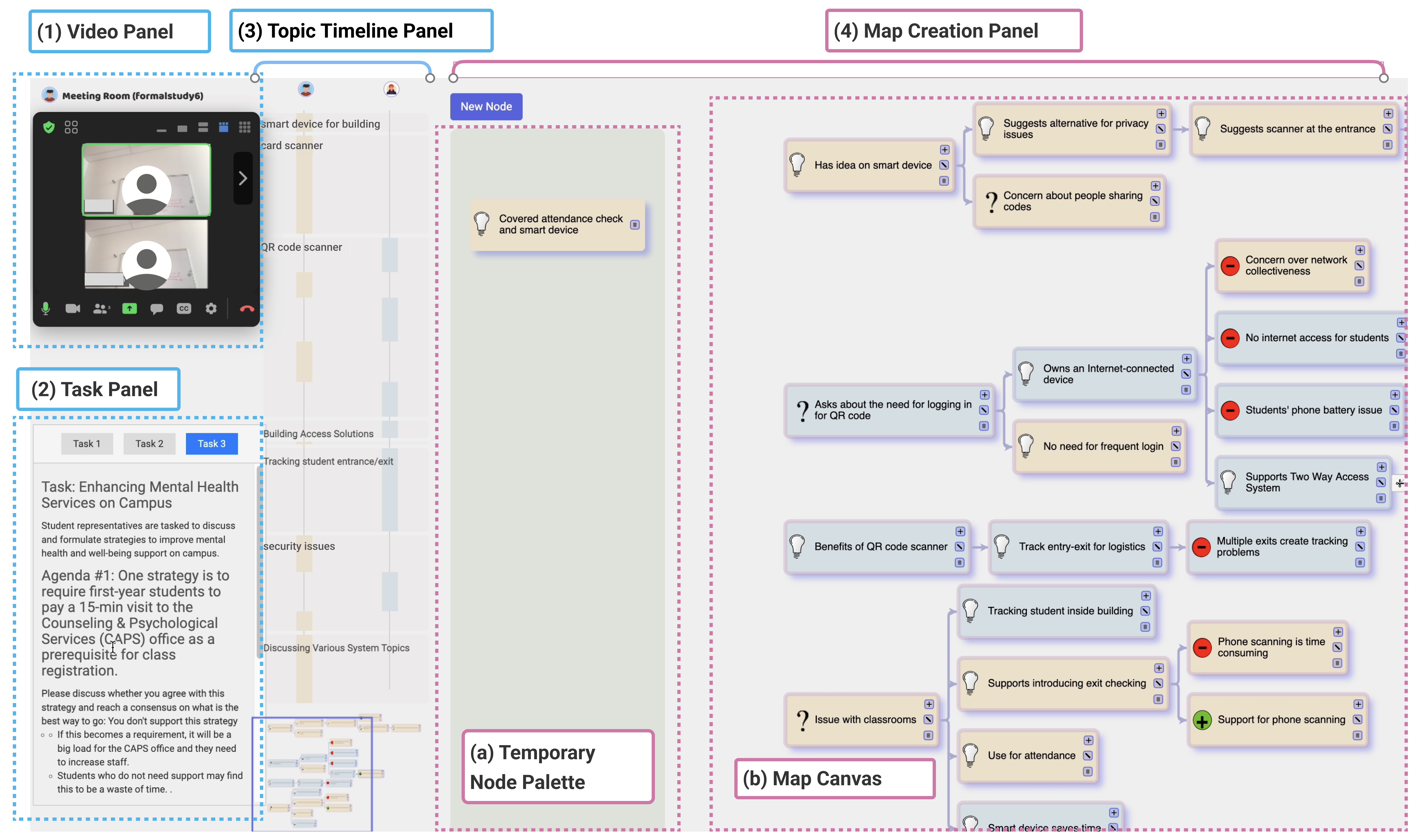}
    \caption{\textbf{MeetMap User Interface}. MeetMap allows users to create dialogue maps collaboratively in real-time during online meetings with LLM support. The system has the following components: (1) a \textsc{Video Panel} and (2) a \textsc{Task Panel}, which displays the meeting agenda; (3) a \textsc{Topic Timeline Panel}, which shows turn exchanges and conversation topics chronologically; (4) a \textsc{Map Creation Panel} that includes two parts: a) A \textsc{Temporary Node Palette} which presents nodes that summarize user speech by AI, arranged in chronological order;  b) A \textsc{Map Canvas} where people can collaboratively create and refine a dialogue map. }
    \label{fig:overview}
\end{teaserfigure}
\maketitle

\section{Introduction}

Keeping up with and making sense of the information exchanged in online meetings is crucial yet challenging \cite{banerjeeNecessityMeetingRecording2005, a.allenUnderstandingWorkplaceMeetings2014, murrayInformationProcessingOverload2019}. 
Turn exchanges happen in rapid succession and progress chronologically, which does not align with the nonlinear way that ideas develop during conversations and how people process and organize information in their minds \cite{clark1991grounding, seuren2021whose}. 
The lack of parallel communication channels and body language cues \cite{yuExploringHowWorkspace2022, chen2023MeetScript}, and the temptation to multitask \cite{cao2021large} further the challenge in online video meetings.

%In contrast to post-meeting systems such as dashboards \cite{samrose2018coco, samrose2021meetingcoach} and auto-generated meeting minutes \cite{asthana2023summaries, hsuehImprovingMeetingSummarization2009}, 
To address these challenges, recent work has provided real-time support for meeting attendees to understand meeting content \cite{chen2023MeetScript, aseniero2020meetcues}. %Real-time sense-making of meeting content is essential to address misunderstanding \cite{clark1987collaborating}, facilitate subsequent participation \cite{yang2022towards}, and build shared understandings between team members \cite{qu2008building, limDesigningSupportSensemaking2021}. 
This includes providing enrichment materials in the context, such as images \cite{liu2023visual} and participation polls \cite{xiaCrossTalkIntelligentSubstrates2023}. %However, there has been limited work on helping people comprehend the conversation in real-time and foster a shared understanding among team members. 
Although these mechanisms can enhance active engagement, they do not help participants fully comprehend conversations in real-time or foster a shared understanding among team members. Some systems have aimed to bridge this gap; for example, TalkTraces maps discussion content to the meeting agenda in real-time\cite{chandrasegaran2019talktraces}, and MeetScript provides real-time transcripts that participants can collaboratively annotate to make sense of meeting content \cite{chen2023MeetScript}. 
However, previous studies have shown that when providing real-time cognitive assistance to meeting participants, it can place a higher cognitive burden on people because they need to consume additional information simultaneously, for example, transcripts and visualizations \cite{chandrasegaran2019talktraces, chen2023MeetScript}.
Additionally, commercial tools have introduced real-time summaries powered by LLMs \cite{Otter.ai_2023}. Yet, studies reveal that these AI-generated summaries can be too abstract and difficult to interpret \cite{asthana2023summaries}, failing to capture the complexity of discussions, particularly the interplay of linear and nonlinear information that is critical during the discussion process \cite{xu-etal-2022-narrate, convertino2008articulating}. 
%Although systems such as TalkTraces and MeetScript promise to provide visualizations and interaction mechanisms for people to make sense of discussions, they also introduce challenges. For example, MeetScript users reported an increased level of cognitive load when managing to speak and read verbose transcripts in real-time \cite{chen2023MeetScript}. TalkTraces users found the topic-centered visualizations to be 
%at times distracting \cite{chandrasegaran2019talktraces}. 

These findings highlight the need for novel interaction and visualization methods to provide real-time sensemaking and comprehension support for meeting attendants while not imposing a cognitive burden on them.
Dialogue mapping \cite{conklin2001facilitated, conklin2005dialogue}  offers a promising solution by visually representing discussions with real-time summaries using a schema.
To practice dialogue mapping in meetings, it requires a designated facilitator who creates and updates the "dialogue map" of the conversation in real-time by capturing the conversation into `questions', `ideas', `pros', and `cons', and linking them into a coherent map; meeting attendees can then communicate with the facilitator to modify the dialogue maps \cite{conklin2005dialogue}.  
In contrast to existing methods that visualize information linearly, such as real-time transcripts and summaries \cite{chandrasegaran2019talktraces, chen2023MeetScript, asthana2023summaries}, the facilitator creates the dialogue maps as the conversation goes, offering a structured visual format that encapsulates the intricate and nonlinear nature of discussions \cite{conklinDialogMappingReflections2003, conklin2001facilitated}. 
% This incremental mapping method enables attendees to follow the discussion dynamically, in contrast to existing transcript \cite{chen2023MeetScript} or summary-based methods\cite{asthana2023summaries}.
However, the requirement of a dedicated facilitator makes this approach impractical \cite{shinChatbotsFacilitatingConsensusBuilding2022, seuren2021whose}. Furthermore, it may remove desirable sense-making opportunities by other participants in the meeting \cite{kiewra1989review}.

To amplify the benefit of dialogue mapping in online video meetings, 
we implement and study the idea of "collaborative dialogue mapping" in video meetings, where meeting attendees can collaboratively build dialogue maps to facilitate their real-time understanding of the conversation. This research addresses the question of when meeting attendees are tasked to create dialogue maps collaboratively during video meetings, how would that influence their discussion quality, understanding of the conversation, and meeting experiences compared to business-as-usual meeting setups?  
% To optimize the collaborative dialogue mapping experience, we further explore the role of AI in facilitating this process.
% Our focus is on finding interaction mechanisms that strike the right balance of AI support during real-time dialogue mapping. This leads us to the core question:  when using AI to facilitate real-time collaborative sense-making tasks—such as creating dialogue maps—how much AI assistance is ideal to maintain user engagement in the cognitive process while reducing the burden of manual map creation?
To further enhance people’s experience with
dialogue mapping, we explore interaction mechanisms that provide the desired level of AI
support when people dialogue map. We also address the question of when using AI to facilitate a real-time collaborative sense-making task, i.e., creating dialogue maps to understand the conversation,
what is the right amount of AI assistance that the users desire?

To address the two research questions, we designed and developed MeetMap (Figure \ref{fig:overview}), an AI-assisted collaborative dialogue mapping system for meeting participants to make sense of the group conversation in real time during online meetings.  
The MeetMap user interface contains 1) \textsc{Video Panel} (Figure \ref{fig:overview}(1)), which connects to a Zoom meeting; 2) \textsc{Task Panel} (Figure \ref{fig:overview}(1)), which displays the meeting agenda;  3) \textsc{Map Creation Panel} (Figure \ref{fig:overview}(3)(4)), where users can collaboratively create dialogue maps.

Meanwhile, MeetMap provides AI assistance to facilitate the creation of dialogue maps. Prior work on human-AI interaction has shown that it is critical to give users control over when and how to use the AI outcomes \cite{amershi2019guidelines, yang2020re}, especially for high-stakes creativity tasks \cite{lu2023readingquizmaker}. To understand how much AI assistance users desire to support sense-making in meetings, we implemented two variants of MeetMap with different levels of human involvement and AI proactivity: Human-Map and AI-Map. During the meeting, participants' speech is transcribed and summarized into nodes by AI, which appear in the \textsc{Temporary Node Palette} following a chronological order. 
In Human-Map, AI generates the summary nodes, and users will create the maps themselves. Users can drag and drop the nodes to the \textsc{Map Canvas} to create dialogue maps. 
In AI-Map, AI generates drafts of dialogue maps, which users can further edit as they wish.

We conducted an evaluation study with 20 participants divided into 10 pairs. Each pair of participants experienced all three conditions, including the two MeetMap variants ( Human-Map and AI-Map) and a baseline condition. We adopted a business-as-usual baseline condition in which participants discussed over Zoom while taking notes collaboratively in a Google Doc with AI-generated summaries.
In each condition, the pair of participants had a 30-minute discussion to make decisions about a task. Here is a summary of the findings.
\begin{enumerate}
    \item Participants found MeetMap (both Human-Map and AI-Map) more helpful than the baseline condition in helping them keep track of the meeting content in real-time and facilitate subsequent discussion. Users liked that MeetMap allowed them to collaboratively organize information through dialogue maps without adding more cognitive burden.
    \item MeetMap helped people build a shared understanding and facilitated more structured subsequent discussions than the baseline. The AI-generated nodes were perceived as objective mediators for people to address misunderstandings.
    \item The presentation of intermediate AI outputs (i.e., the AI-generated nodes) was perceived to enhance the system's synchronicity. Users felt they could engage with the AI-generated nodes in real-time, whereas in the baseline condition, the periodic summaries lacked sufficient synchrony.
    \item Users found it easier to refine AI-generated content than to edit peer-created notes, as modifying AI-generated content felt less intrusive. This fostered a more collaborative environment, leading to more collaboration in creating shared notes in the MeetMap conditions.
    \item Between the two MeetMap variants (Human-Map and AI-Map), users reported feeling greater agency and control when creating their own dialogue maps in Human-Map. Participants in Human-Map noted that the process of actively constructing the maps enhanced their understanding of the conversations, in contrast to AI-Map, where they passively received information. Users appreciated the more intensive cognitive engagement offered by Human-Map when seeking to make sense of the content in real time. In contrast, they valued AI-Map's ability to generate a structured output that facilitated post-meeting reviews. 
    \item Users experienced a stronger sense of ownership over the resulting dialogue maps in Human-Map. They displayed lower tolerance for AI errors when they perceived themselves as responsible for the final deliverable. 
\end{enumerate}
%  (as shown in \ref{cognitive}). Meanwhile,  the AI-generated nodes in MeetMap were perceived as objective mediators to address misunderstandings. 
% Second, users in Human-Map considered the AI-generated content to be synchronous and succinct so that they could read and act on it in real time. Furthermore, users wanted to participate in cognitive and creative thinking processes in Human-Map, even if it required more work.
% Our findings indicate that MeetMap enhances discussion quality and improves users’ understanding of conversations compared to the baseline (5.1). Both the Human-Map and AI-Map helped individuals keep up with the discussion (5.1.1) and facilitated team consensus-building and collaborative sense-making, with AI-generated nodes viewed as  neutral mediators to address misunderstandings (5.1.2). MeetMap also encouraged balanced and equitable engagement in note-taking  without increasing cognitive load (5.1.3, 5.1.4). Our findings also reveal that varying levels of AI assistance influence user engagement and experience in distinct ways (5.2). The Human-Map encouraged greater user agency and deeper cognitive involvement (5.2.1), while the AI-Map simplified map creation through pre-structured outputs and facilitated post-meeting review (5.2.2). Users expressed a preference for engaging in cognitively and creatively demanding processes, even when these required greater effort (5.2.3). However, users showed lower tolerance for AI errors when they perceived ownership of the human-AI collaborative output (5.2.4).

Through designing and developing MeetMap, we summarize design strategies for developing LLM-powered systems to facilitate real-time collaboration. 1) Delays in using AI to synthesize conversations (e.g., generating summaries and dialogue maps) can affect user experience. To address this, presenting intermediate AI outputs can enhance the system's perceived synchrony and transparency. In MeetMap, the introduction of the Temporary Code Palette was widely praised by users, as it displayed summary nodes in real-time, allowing them to gradually understand the content before the full dialogue map was generated. 2) Enhancing users' cognitive engagement is particularly beneficial for cognitively demanding tasks, such as collaborative conversations. While Human-Map offers less automation compared to AI-Map, it was praised by users for providing more opportunities to actively engage with and make sense of the content.
3) Using visualizations to enhance the structure of AI-generated content can reduce users' cognitive load, especially when they need to process information in real time. In MeetMap, users appreciated features such as color-coding for different speakers and the dialogue map notation schema, which categorized nodes into four categories.

\section{Related Work}

\subsection{Challenges of Keeping up with and Making Sense of Online Meetings }
Group meetings are events where participants discuss, negotiate, present, and create materials together in a communicative manner \cite{straus1996getting}. To make the group meeting effective, it is critical for meeting participants to keep up with the ongoing conversations \cite{echenique2014effects}, make sense of the discourse \cite{limDesigningSupportSensemaking2021, telenius2016sensemaking}, and foster a shared understanding \cite{brennan1998grounding, birnholtz2005grounding, clark1989contributing}. 

However, significant challenges arise in real-time comprehension and sense-making of discussions. One primary reason is a lack of visual representation of conversations, making it hard for people to revisit missed content \cite{shi2018meetingvis, banerjeeNecessityMeetingRecording2005}. Moreover, interpreting and making sense of conversations involves an associative memory process, where individuals link scattered topics over time to form a cohesive understanding, while people's working memory and attention spans are limited \cite{wangIdeaExpanderSupporting2010}. The challenges become more severe when the session is running long \cite{karl2021virtual}, the meeting is poorly structured \cite{karl2021virtual}, and the discussion is back-and-forth\cite{brennan1998grounding, convertino2008articulating} with irrelevant side discussions \cite{hou2011analyzing}.
Additionally, sense-making in meetings is not only an individual behavior but a collective move, wherein meanings are progressively built through collaboration \cite{qu2008building}. This process of grounding, or building a shared understanding, is central to successful collaboration \cite{clark1991grounding}. However, building a shared understanding among team members can be extra challenging for virtual or hybrid meetings due to the lack of physical shared space \cite{kirshenbaumTracesTimeSpace2021c, yuExploringHowWorkspace2022}, social cues \cite{murali2021affectivespotlight}, and technological barriers and fatigues \cite{bailensonNonverbalOverloadTheoretical2021}.

\subsection{Technologies to Support Sense-making in Online Group Meetings}
Researchers have explored various technological interventions to help people keep track of and make sense of synchronous group meetings. The main focus is post-meeting tools, such as dashboards with visualizations designed to help users review content and reflect on their participation \cite{samrose2018coco,samrose2021meetingcoach}. However, these tools fall short of aiding real-time sense-making. We have witnessed more research in recent years to support real-time sense-making and understanding of meeting content. 
One line of work visualizes non-verbal cues to mimic the communication experience in face-to-face discussion \cite{shi2018meetingvis, aseniero2020meetcues, leshedVisualizingRealtimeLanguagebased2009, murali2021affectivespotlight}, for instance, amplifying expressions and body language to enhance situational awareness \cite{aseniero2020meetcues, murali2021affectivespotlight, aseniero2020meetcues}. Another line of work uses real-time dialogue analysis and visualization to improve situated understanding \cite{chandrasegaran2019talktraces, junuzovic2011did,  liu2023visual}. TalkTraces, specifically, links current discussions with past meeting agendas \cite{chandrasegaran2019talktraces}. Additionally,  web-based collaborative whiteboards have been used for sharing screens and discussion resources, creating external memory aids and visual representation of the conversation to reduce cognitive load \cite{yuExploringHowWorkspace2022}.
However, these approaches often present information passively, missing opportunities for active user engagement in sense-making, which is key for understanding and memory \cite{schwartz2016abcs}. MeetScript introduced collaborative annotation of live transcripts, showing that actively engaging users to make sense of the transcripts aids comprehension \cite{chen2023MeetScript}. But the verbosity of full transcripts can be cognitively burdensome \cite{fang2021notecostruct, chen2023MeetScript}, and the linear presentation of information is misaligned with the iterative and structural nature of how people understand the conversation \cite{clark1991grounding, seuren2021whose}.

Recognizing these gaps, this paper presents MeetMap to aid individuals in understanding meeting content in-situ. First, we aim to actively engage users in the sense-making process, supporting them in creating an external representation of the discussion collaboratively. Second, we aim to reduce the unnecessary cognitive load on processing real-time comprehension aids, such as verbatim transcripts. Third, in contrast with the linear representation of the conversation following a chronological order,  we aim to provide more flexibility for users to organize their thoughts.

\subsection{Collaborative Note-Taking and Concept-Mapping Supports Sense-making in Teams} 

Collaborative note-taking enables meeting participants to create shared notes \cite{Barnett1981WhatIL, Jansen2017AnIR, mik2019effects, costley2021collaborative}, which is found to enable real-time clarifications, reduce misinterpretations, and help the team build a shared understanding \cite{davis1998notepals, richter2001integrating, kam2005livenotes} in comparison to individual note-taking. Moreover, collaborative note-taking reduces the burden on individual team members \cite{kam2005livenotes, fanguy2023analyzing, makany2009optimising} and improves the quality of the shared notes with diverse viewpoints \cite{dyke2013student}.

Collaborative note-taking methods can be categorized into note-taking following a chronological order and note-taking non-linearly and visually \cite{davis1998notepals, richter2001integrating, fanguy2023analyzing, makany2009optimising, kam2005livenotes}.
Studies have shown that taking collaborative notes chronologically enhances engagement and reduces distraction \cite{fang2022understanding}
, however, it also can be insufficient to support sense-making for complex information \cite{cao2022videosticker}. On the other hand, taking notes visually and non-linearly allows for flexibility and better integration of ideas through various forms of external representations like concept maps and diagrams \cite{shih2009groupmind, liu2018conceptscape}.

Prior work has shown that creating concept maps collaboratively as a team improves learning of lecture videos \cite{liu2018conceptscape, cao2022videosticker} and brainstorming of new ideas \cite{sun2022students, shih2009groupmind, lansiquot2015concept}. %}.
%It has also been shown that students can better regulate their conversations and expand ideas when they create digital concept maps during the group discussion \cite{sun2022students, 
A similar idea, called ``Dialogue Mapping'', has been explored in n-person group meetings, which uses a concept map to help teams visualize the progression of ideas in real-time, serving as a team facilitation technique\cite{conklin2005dialogue}.   Mapping categorizes conversations into nodes with IBIS notation schemas, including Questions, Ideas, and Pros and Cons arguments, and visually connects these nodes to form a 'map' \cite{conklin2001facilitated}.
Previous dialogue mapping practices often require a dedicated facilitator to create the dialogue map during the meeting \cite{conklinDialogMappingReflections2003}, limiting the practical use of this approach. 
Moreover, employing a dedicated facilitator also takes away the cognitive processes people may engage in to make sense of the meeting content actively \cite{kiewra1989review}. Additionally, creating a high-quality dialogue map demands substantial manual effort, making it particularly challenging in synchronous communication scenarios like video meetings, which are already cognitively demanding \cite{dhawan2021videoconferencing}.
Building upon prior work,  we aim to integrate collaborative dialogue mapping capabilities in video meeting contexts. More specifically, we aim to redistribute the labor of creating dialogue maps to all meeting participants and use AI to scaffold the process of creating dialogue maps as a group.

\subsection{The Use of AI to Support Real-Time Communication}
With the recent advancement of large language models (LLMs), researchers have begun to leverage them to support sense-making of texts, e.g.,  facilitating text comprehension with LLM-generated diagrams \cite{Jiang2023GraphologueEL, Suh2023SensecapeEM}, and assisting reading and writing \cite{lu2023readingquizmaker, kangSynergiMixedInitiativeSystem2023, zhangVISARHumanAIArgumentative2023}. 
Using LLMs to structure real-time spoken conversations introduces extra challenges \cite{liangImplicitCommunicationActionable2019, hancock2020ai}. 
It places a higher emphasis on presenting the essential information to users while they manage to speak, listen, and comprehend the discussion content simultaneously \cite{asthana2023summaries, chen2023MeetScript}.

Prior research has seen initial successes using AI to enhance human-human communication \cite{liangImplicitCommunicationActionable2019, hancock2020ai}. For example, real-time transcripts and translations help cross-cultural discussion \cite{zhangFacilitatingGlobalTeam2022}, and conversational agents can facilitate communication and collaborative decision-making \cite{kim2020bot, kim2021moderator}. 
However, similar to using AI in other activities that require cognitive engagement, using AI to facilitate in-situ sense-making in conversations may introduce an assistance dilemma \cite{koedinger2007exploring}. For example, early studies have shown that when one team member is responsible for documenting the content of a meeting, it limits the other attendees' engagement with the content \cite{reilly2011groupnotes}. In contrast, engaging all meeting participants in note-taking reinforces their understanding and retention \cite{kam2005livenotes, mahyar2012note}. %Thus, careful design is required when providing AI assistance in collaborative sense-making activities to not take away people's opportunities to actively process the content. 
Indeed, it is shown that when using AI to support human-human communication, over-reliance on AI may occur \cite{asthana2023summaries}. It discourages people from actively engaging in the conversation and may weaken people's skills in regulating and organizing the conversation \cite{kim2021moderator}. Prior work has argued that AI should act as an enabler and not a replacement in human communication \cite{zheng2023competent}, users should have more control over when and how to use information provided by AI \cite{liu2023visual}, and that it is critical to provide just the right amount of AI-generated content to users \cite{sonItOkayBe2023, amershi2019guidelines}. Our work closely follows the recommendations made in the literature on AI-mediated communication, in which we aim to give users agency and flexibility in organizing their thoughts.

\textbf{As a summary of previous work}, people continue to face challenges in keeping track of and actively making sense of discussions in real-time. Dialogue mapping has been shown to be a successful meeting facilitation technique \cite{conklin2005dialogue}. However, the requirement of designated facilitators limits the practical use of the approach. This work aims to integrate collaborative dialogue mapping capabilities in video meetings to enhance people's real-time understanding of the conversation. To facilitate the dialogue mapping process, we provide people with AI assistance. We also build upon prior work on AI-mediated communication to ensure that AI provides users the right amount of assistance.

\section{MeetMap: Generating Dialogue Maps as Real-Time Cognitive Scaffolds for Online Meetings}

\subsection{Design Goals} \label{design-goal}
% Informed by the previous work,  we identified four design goals for employing AI to aid real-time sense-making through Dialogue Mappings in meetings.
We propose MeetMap, an AI-assisted collaborative dialogue mapping system for meeting participants to make sense of the group conversation in real-time during the meetings. The design of MeetMap is inspired by prior work on dialogue mapping \cite{conklin2005dialogue} and collaborative sense-making tools \cite{chen2023MeetScript, liu2018conceptscape, zhang2018making}.
MeetMap is designed to support collaborative sense-making in a new context that has not been explored before, namely for meeting attendees to develop a shared understanding of the meeting content in situ with the collaborative dialogue map creation. Below, we list the design goals derived from the prior work and solutions we employed.

\begin{itemize}
    \item \textbf{Design Goal 1: Enable people to collaboratively build dialogue maps (D1).} Existing dialogue mapping techniques have been found to enhance real-time understanding by employing a notation schema and visual structure \cite{conklin2001facilitated, conklin2005dialogue}. %In MeetMap, we adopted this visual structure and the notation scheme to support real-time meeting sense-making.   
    However, a significant limitation of current dialogue mapping techniques is the need for designated facilitators \cite{zubizarreta2013co, ng2008dialogue}, which might take away the beneficial cognitive processes when people engage in such activities actively \cite{koedinger2007exploring}. Prior work on collaborative sense-making highlights the importance of active engagement through creating external representations collaboratively rather than passively receiving information \cite{zhang2018making, zhang2017wikum}. Inspired by these insights, MeetMap provides a shared space where meeting participants can collaboratively create and edit the dialogue maps as the conversation unfolds.
    
    \item \textbf{Design Goal 2: Help people manage the cognitive load of creating dialogue maps during the conversation (D2).} Creating dialogue maps can be a cognitively demanding process due to the complex visual structure it requires \cite{conklin2005dialogue, piolat2005cognitive}. While encouraging users to collaboratively create dialogue maps, the system should also reduce the extra cognitive effort brought by the collaborative map creation. %with AI assistance
     To achieve this, the design of MeetMap leverages AI to categorize turns and generate short summaries of conversations. It enables users to have the capacity to peruse these AI-generated summaries in the creation of dialogue maps. 
    
    \item \textbf{Design Goal 3: Help people easily make sense of the AI assistance when creating dialogue maps (D3)}. Prior work on using AI to support collaborative sense-making informs the need to present information at varying levels of detail, from overarching themes to specific utterances \cite{chandrasegaran2019talktraces, xiaCrossTalkIntelligentSubstrates2023}. Prior work also suggests supporting seamless interaction between linear and nonlinear representation to avoid losing context when understanding diagram solely \cite{jiangGraphologueExploringLarge2023}. Taking inspiration from those studies, we devised several visualization techniques to assist users in interpreting AI-generated content in the development of MeetMap. This includes 1) showing the original transcript of the summaries so that people know where the summaries come from; %2) displaying summaries promptly based on short conversation chunks to mitigate delay; 
    2) differentiating the turns by different speakers; 3) organizing the discussion in chronological sequence with dialogue maps to help participants associate linear and nonlinear information during meetings.
    
    \item \textbf{Design Goal 4: Identify the right level of AI assistance to provide (D4).} Providing more AI assistance may limit the opportunity for users to engage actively in the sense-making process \cite{liao2023deepthinkingmap, dhillon2024shaping}.  Extensive research in human-AI collaboration emphasizes the importance of maintaining user agency \cite{zhangVISARHumanAIArgumentative2023, amershi2019guidelines, zhang2021ideal}.  Drawing insights from these studies, we aim to design dialogue mapping systems that seek a desired level of AI assistance with human involvement and control.  In the design of MeetMap, we employ two levels of AI assistance. In \HumanMap, AI generates short summaries where users will create dialogue maps leveraging the summaries. In \AIMap, AI directly outputs dialogue maps. 
    
\end{itemize}

\subsection{Iterative Design and Development} 
The goal of MeetMap's design is to develop an AI-assisted collaborative dialogue mapping system that aids meeting participants in understanding conversations in real-time. Nonetheless, the optimal degree of AI assistance and the method of information presentation to reduce cognitive load have yet to be determined.
The design and development of MeetMap followed an iterative, user-centered approach, refined through several pilot tests and feedback cycles to optimize usability, cognitive load management, and real-time collaborative mapping with AI assistance. This section provides a detailed explanation of these iterations and the insights gained at each stage.

\subsubsection{Overview of the initial design}
The initial version of MeetMap was designed to support collaborative dialogue mapping in real-time online meetings, addressing the four core design goals as mentioned in section \ref{design-goal}. 

The core design of MeetMap’s features has remained consistent from the initial to the final version. The core function of MeetMap contains a Map Creation Pane,  including a Temporary Node Palette with AI-generated summary nodes and a Map Canvas where users collaboratively build a conversation map, as shown in Figure \ref{fig:overview}. This design enables active engagement in shared mapping \textbf{(D1)} while managing cognitive load \textbf{(D2)} by generating categorized summaries in near real-time. Visual aids, including color-coded speakers and a Topic Timeline, help users link AI content with the conversation’s flow \textbf{(D3)}. Additionally, the system provides two levels of AI assistance: in \HumanMap, users build the map from AI-generated nodes to structure their ideas, while in \AIMap, the system drafts a structured map for users to refine, offering different levels of guidance \textbf{(D4)}. Detailed description is shown in section \ref{systemdesign}. However, the early version of \AIMap initially appended new nodes to create a single, comprehensive map covering the entire conversation—similar to facilitator-led dialogue mapping, where one person continuously adds to a complete conversation map.  %Users also found it hard to understand the chronological flow of the conversation with the visual structure on the map. To resolve this, when users click the topic in the \textsc{Topic Timeline Panel}, the nodes on the map are highlighted to help people connect the linear conversation with the map.

\subsubsection{Pilot-test and Improvement}
We conducted three pilot tests (3 groups of 2) with the early version of MeetMap. During these pilots, participants completed group discussion tasks similar to the formal studies (detailed in Section 4). After each discussion, we interviewed participants for feedback on the system's usability.

Some detailed design decisions are made based on user feedback.
\textbf{(1) First}, feedback indicated challenges with real-time node updates, as the delayed appearance of nodes interrupted the conversation flow. In response, we implemented a turn-splitting mechanism, as detailed in section \ref{split}, to display AI-generated nodes immediately after each turn, ensuring participants could access summaries in near real-time without waiting for delayed processing.\textbf{ (2) Second,} the early version of \AIMap automatically generated a large dialogue map for the entire conversation, which some users found overwhelming. Users preferred to receive information gradually to help them digest it. Based on this feedback, we transitioned to a topic-based segmentation, where smaller, digestible maps are generated incrementally, as detailed in section \ref{AI-Map}. 
\textbf{(3) Third, }as maps grew larger, users encountered navigation challenges. To address this, we introduced interactive connections between the  \textsc{Map Canvas}, \textsc{Temporary Node Palette}, and \textsc{Topic Timeline}, enabling users to seamlessly navigate between linear and non-linear information, as shown in section \ref{interaction}. Additionally, we incorporated a mini-map at the bottom of the screen, which makes navigation more intuitive.

%The final system was re-tested to ensure smooth user interaction and to confirm there were no critical usability issues before the formal study.

\subsection{System Design}
\label{systemdesign}
% \xy{2AC: There is a disconnect between the design challenges (page#6) and the system design (section 3.1). The system design should discuss what measures were taken or what components were added to address each of the challenges. Explicitly addressing them will also make the system design decisions more powerful.
% }

 We reported the final system design of MeetMap. 
 An overview of the MeetMap system is shown in Figure \ref{fig:overview}. MeetMap's front-end interface has three components: 1) A \textsc{Video Panel} for users to have online meetings (Figure \ref{fig:overview}(1)); 2) A \textsc{Task Panel} to upload meeting agendas (Figure \ref{fig:overview}(2)); %3) A \textsc{Topic Timeline Panel} to show the turn-taking dynamics and the general topics discussed following a chronological order (Figure \ref{fig:overview}(3)); 
3) A \textsc{Map Creation Panel} which contains a \textsc{Temporary Node Palette} (Figure \ref{fig:overview}(a))and \textsc{Map Canvas}(Figure \ref{fig:overview}(b)), which enable users to create a shared representation of meeting content in real-time.

\subsubsection{Map creation panel}
MeetMap offers a \textsc{Map Creation Panel} that allows users to collaboratively and flexibly construct dialogue maps during online meetings with ease \textbf{(D1)} while also being designed to prevent overwhelming users with the additional content supplied by AI \textbf{(D2)}.

\begin{figure}[h]
    \centering
    \includegraphics[width=\textwidth]{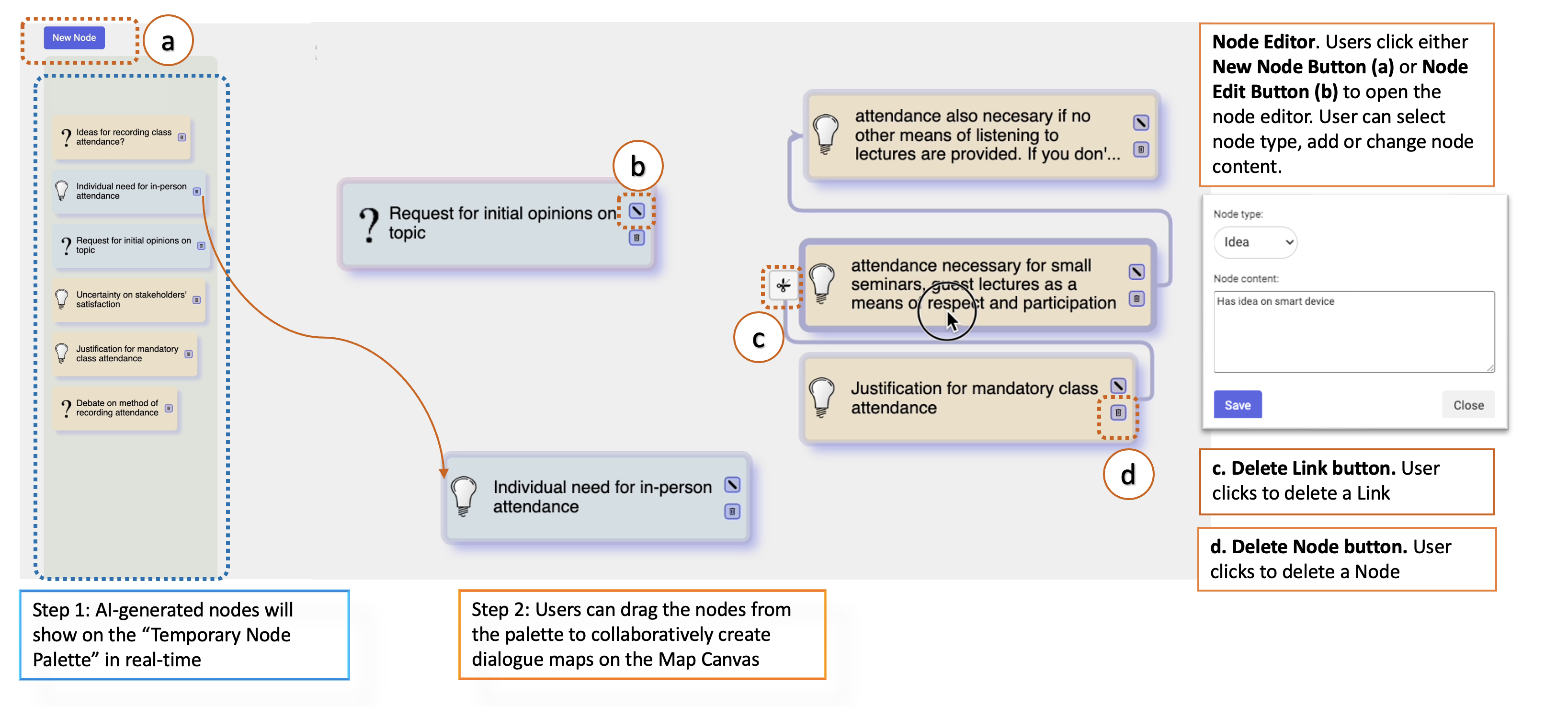}
    % \caption{The workflow of how users interact in \HumanMap.}
    \caption{\textbf{Users can collaboratively create dialogue maps in MeetMap.} Users can use the AI-generated nodes to create the map. (1) Nodes are shown in \textsc{Temporary Node Palette} in real-time. (2) Users can drag nodes to the map and create links between the nodes. Using the interaction suite (a-d), users can create/delete nodes (a, d), edit nodes (b), and delete links (b). When creating/editing a node (b), users can specify the node content and node type.}
    \label{fig:Map-creation}
\end{figure}

\textbf{Map Canvas.}
 MeetMap provided a suite of map interaction features, illustrated in Figure \ref{fig:Map-creation}, to enable users to create dialogue maps collaboratively \textbf{(D1)}. As supported by other collaborative concept map authoring tools, e.g., Miro Board, users can collaboratively create, edit, and delete nodes and links on the \textsc{Map Canvas}. These features were provided to meet the basic needs of creating a collaborative dialogue map of group meetings. In addition to editing the node content, users can assign a node category using the "dialogue mapping" notation schema \cite{conklin2005dialogue}.

\textbf{Temporary Node Palette. } \label{split} To help people manage the cognitive load of creating dialogue maps during the conversation \textbf{(D2)},  MeetMap displays AI-generated summaries of users' conversations as nodes on the \textsc{Temporary Node Palette} (Figure \ref{fig:Map-creation} (1)) to reduce the human effort of creating notes.
MeetMap detects each turn as a single unit to generate the summary. The system recognizes speakers, detects and transcribes turns in real-time using the Azure speech-to-text service \footnote{https://azure.microsoft.com/en-us/products/ai-services/speech-to-text}. Since participants typically take notes on what has just been discussed, the AI-generated summary should be shown almost synchronously \cite{bothinUserEvaluationStudy2014}. To mitigate the lag from lengthy monologues, MeetMap employs a 50-word checkpoint to prompt new conversation segments. This threshold, informed by studies indicating an average turn length of 50 words in group meetings \cite{chen2023MeetScript}, does not abruptly end conversations at 50 words but waits for a natural sentence closure. Azure's speech-to-text service detects this natural sentence end. If a sentence concludes after surpassing 50 words, the system initiates a new turn, even without a change of speaker. This strategy prevents both mid-sentence interruptions and reduces delays from extended monologues.

\begin{wrapfigure}{l}{0.5\textwidth}
\includegraphics[width=\linewidth]{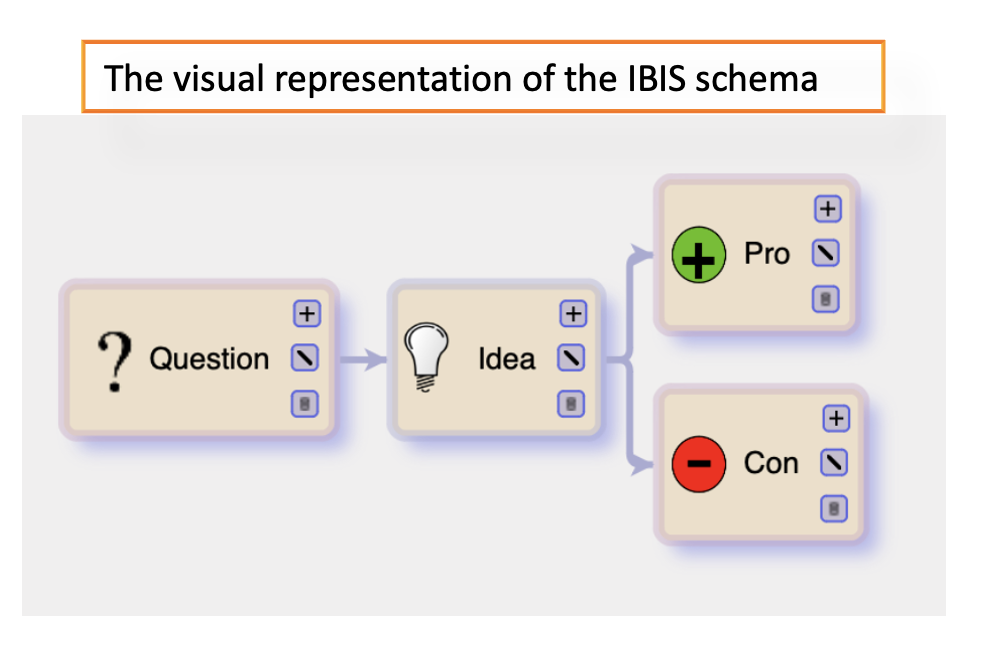} 
\caption{The visual representation of the IBIS notation schema: We use symbols to visualize the dialogue mapping notation schema, including "Questions, ideas, pros, and cons"}
\label{fig:notation}
\end{wrapfigure}

For each detected turn, it is sent to GPT4 to 1) assign a tag based on the IBIS notation schema used in dialogue mapping techniques\cite{conklin2005dialogue}, as shown in Figure \ref{fig:notation}. The four categories of the IBIS notation schema include \textbf{Questions}: problems or issues that need to be addressed; \textbf{Ideas}: responses and proposed solutions to issues; Arguments includes \textbf{``Pro"} and \textbf{``Con"}, as justifications supportive or against a particular idea. 2) generate summary nodes of this turn. The AI will generate one or several summary nodes in one turn if the different parts in one turn belong to various categories.   
We used the GPT-4 API to categorize turns and generate summaries. % in our system as previous studies have shown the remarkable performance of GPT-4 in summarization \cite{xu-etal-2022-narrate} and zero-shot learning methods for text classification \cite{tornberg2023chatgpt}.  
The full prompt used in the system can be found in the Appendix. \ref{appendix:prompts}.  The output will appear as one or several nodes in the \textsc{Temporary Node Palette} in real time following a chronological order, as shown in Figure \ref{fig:Map-creation} (1). 
Users can delete any node from the \textsc{Temporary Node Palette}. Our system allows no minimum turn length, meaning concise responses like "Agreed" or "Yeah!" are evaluated by GPT-4 to assign a dialogue mapping tag with a consistent prompt to maintain tagging uniformity. Non-qualifying turns are excluded and do not produce nodes. This method captures all elements of the dialogue, ensuring that even minor interjections or queries contributing to the conversation’s flow are accurately represented using the IBIS notation schema.

%The IBIS notation schema used in dialogue mapping is simple, intuitive, and powerful enough to capture discussions and the rationale behind decisions \cite{conklin2005dialogue}. 
Years of experience have shown that IBIS does not introduce too much cognitive overhead but has just enough structure to
capture a conversation using only four categories and a limited number of building blocks \cite{isterdaelDialogueMappingGuideaMaps, conklinDialogMappingReflections2003, conklin2001facilitated, conklin2005dialogue}. Although the structure is simple, it effectively promotes asking the right questions and keeping discussions on track % helping to reach conclusions efficiently
\cite{isterdaelDialogueMappingGuideaMaps, wegerifExploringCreativeThinking2010}.  
%The notation schema in dialogue mapping has proved simple and useful for team meetings, thus requiring minimal cognitive load on the users to apply it. %Moreover, the dialogue mapping technique aligns well with the design goals derived from our formative study in that it provides a visual representation and enables users to easily adjust the structure. 
We acknowledge the limitations of the notation schema.  In particular, the schema may not cover the whole range of conversational nuances or be appropriate for all meeting categories. However, the primary goal of using this schema in MeetMap is not to assert that it is universally applicable but to offer it as an example of how AI assistance may aid group understanding through dialogue mapping techniques. MeetMap's other interaction design is agnostic to the notation schema, which allows for the adaptation or replacement of the notation schema if the alternative schema proves more effective in other meeting scenarios. We will further discuss the generalizability of using an AI-generated notation schema to categorize meetings in the discussion.
% if they find it irrelevant. We emphasize the visual representation of the IBIS tag on the nodes based on our user's needs to see more visual information (D3). We adapted the visual symbols of each IBIS tag in our system and made it evident on every node, as shown in Figure \ref{fig:visual} (1).

\textbf{Interaction between the Map Canvas and the Temporary Node Palette.} \label{interaction}
Different from other tools, the interaction of \textsc{Map Canvas} in MeetMap was specifically designed to reduce the effort of creating dialogue mapping collaboratively during meetings and help users to make use of the AI assistance \textbf{(D2)}. The \textsc{Map Canvas} is seamlessly interconnected with the \textsc{Temporary Node Palette} that users can drag the nodes generated by AI in the \textsc{Temporary Node Palette} to create dialogue maps on the \textsc{Map Canvas}.  % Besides editing the node content, users can also specify a category of the "dialogue mapping" schema for each node, and this can support users to categorize conversation turns when the AI-assigned categorize the conversation wrongly or when they manually add new nodes to the shared notes. 

\begin{figure}[h]

\centering
\includegraphics[width=\textwidth]{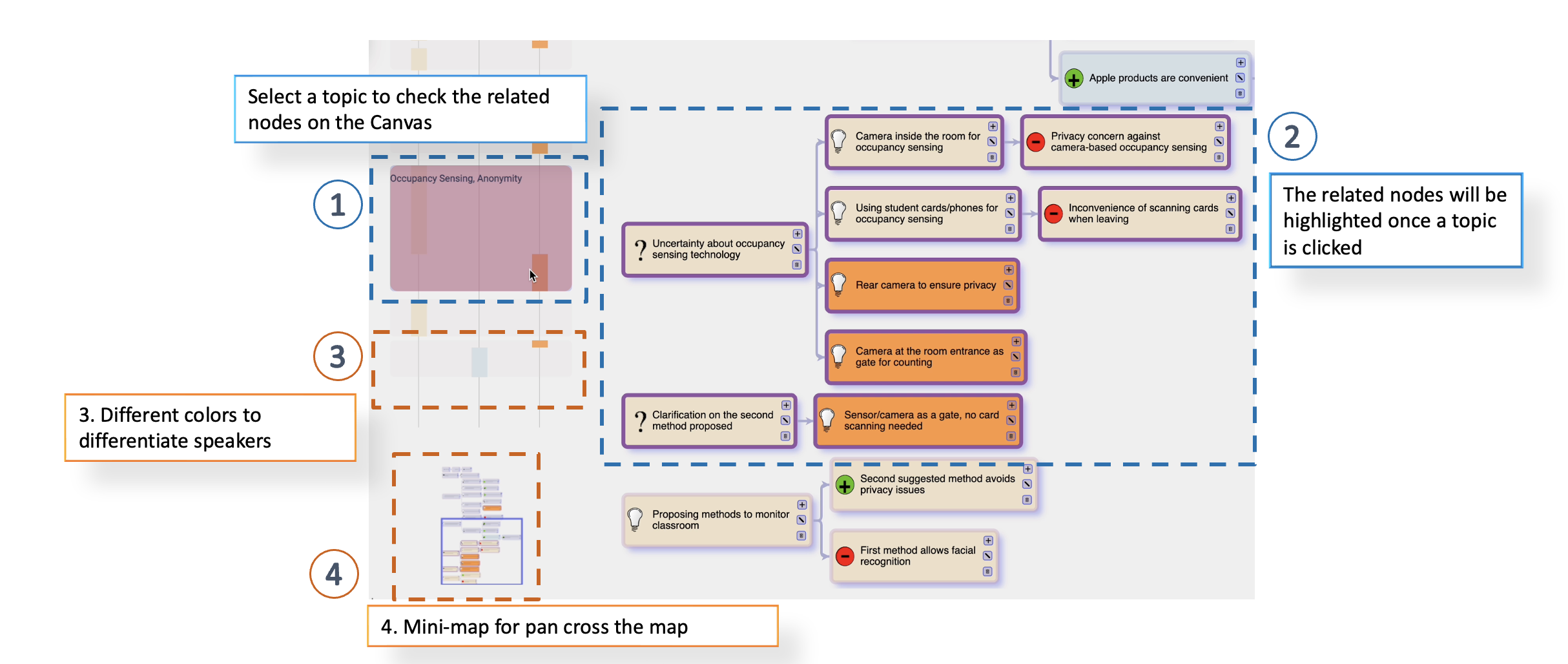}
\caption{\textbf{We introduced a series of visualizations to improve the usability of MeetMap}: (1) The discussion topics are segmented and labeled in the timeline; (2) When users click a topic in the timeline, the topic block changes colors, and the related nodes will be highlighted on the \textsc{Map Canvas}. (3) The timeline view represents speakers in different colors;  (4) Mini-map for map navigation.}
\label{fig:visual}
\end{figure}

% \subsubsection{Topic Timeline Panel}

\subsubsection{Visualizing information with different detail levels}
MeetMap visualizes conversation information with different granularity to help people easily make sense of the AI-generated content when creating dialogue maps \textbf{(D3)}, as illustrated in Figure \ref{fig:visual}. MeetMap summarizes the conversation into summary nodes in \textsc{Temporary Node Palette}. Users can view the original transcript that AI used to generate the node through double-clicking a node. To help people quickly gain an overview of the conversation, MeetMap visualizes the turn-taking and collaborative dynamics through \textsc{Topic Timeline Panel}, with speakers represented in different colors, as shown in Figure \ref{fig:visual}(3).   The AI-generated nodes in the \textsc{Temporary Node Palette} and \textsc{Map Canvas} follow the same color coding for each speaker to visualize individual contributions.   To further meet the needs of connecting chronological information with the visual representation of conversation \textbf{(D3)}, we support users to map information from the \textsc{Topic Timeline Panel} to the \textsc{Map Canvas}.  Users can click on a topic block on the timeline to view the highlighted corresponding nodes on the map canvas (Figure \ref{fig:visual}(1)(2). % In general, the verbal conversation is represented in MeetMap with a visual and color-coded overview, a turn summary with notation schema, and a full textual transcript, allowing people to easily navigate the context and expand to different levels of detail as they desire.  
Additionally, we added a Mini-map at the bottom of the screen. The mini-map is a scaled-down version of the full map, which shows the overall structure and the color-coding of the full map without details, serving as a map locator for moving large canvases when there are many nodes and connections (Figure \ref{fig:visual} (4)).

% For \AIMap, the introduction of the temporary code palette also helped to mitigate the delays on map creation. Moreover, 

% e.g., smaller dialogue maps or AI-generated nodes with summaries shown in real-time. Based on the user feedback, we decided not to generate comprehensive dialogue maps. Instead, we decided to use smaller chunks of conversations as input and generate content that can be understood and used in real-time. 

% \textbf{Second,} some users found that a dialogue map updated only for large conversation chunks (e.g., when topic changed)  may not facilitate active engagement to build the dialogue map collaboratively.  As the "desirable difficulty" theory suggests \cite{bjork2020desirable}, AI takes all the note-taking tasks that may hinder users in experiencing the cognitive challenge that is beneficial to understand. In our system design, we want to ensure users have opportunities and are motivated to make the dialogue map collaboratively with some AI assistance. However, since we are still not sure how much AI assistance is desirable by users, we designed two variants for different levels of AI assistance. In both variants, we maintained the user agency of editing and refine the map collaboratively. Meanwhile, we provided different levels of AI assistance, which means users need pay more or less effort to build the dialogue map by their own, to explore how AI should be introduced to assist the understanding and sense-making in synchronous human communication.

\subsection{Two Variants of MeetMap: \HumanMap and \AIMap}
Informed by both the prior literature on user agency in human-AI interaction \cite{lu2023readingquizmaker, amershi2019guidelines} and AI-mediated communication \cite{kadoma2024role} and the pilot study with the early system design, we designed two variants of MeetMap with different levels of user involvement and AI assistance \textbf{(D4)} %, and the workflow of the two variants are illustrated in Figure \ref{fig:workflow}. 
In \HumanMap, users see AI-generated nodes and will create dialogue maps themselves. In \AIMap, users see AI-generated small dialogue maps automatically, and users can refine them. %The workflow on using AI to generate nodes and maps is shown 

\begin{figure}[htbp]
    \centering
    \includegraphics[width=\textwidth]{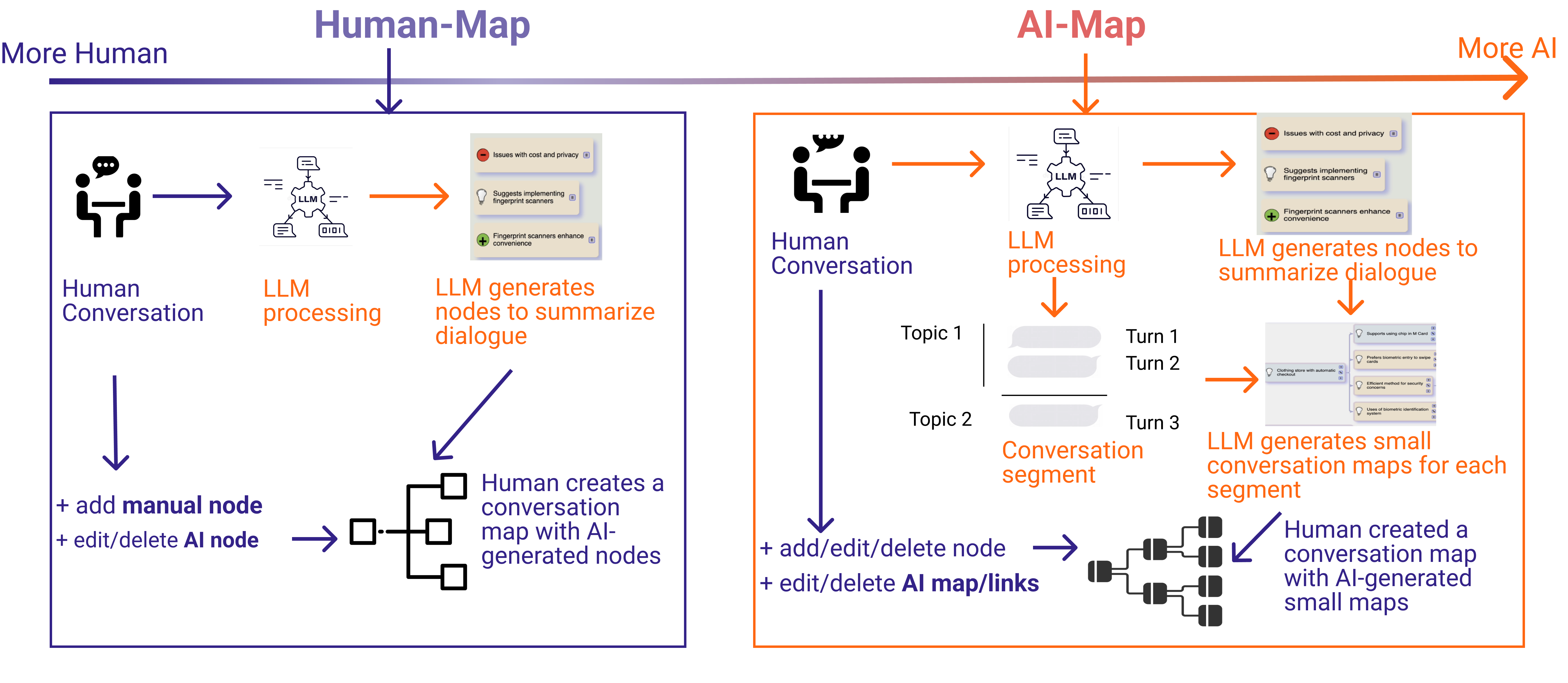}
    \caption{\textbf{Two Variants of the MeetMap System with different levels of human involvement and AI assistance.} In \HumanMap, AI only generates the nodes, and users will create the maps themselves. In \AIMap, AI generates drafts of dialogue maps, where users can further make edits in any way they want.}
    \label{fig:workflow}
\end{figure}

\subsubsection{\HumanMap: AI generates only summary nodes, and humans create the links.}
\HumanMap enables users to create a conversation map that meets their needs. AI will only generate nodes with the dialogue mapping notation schema and present those nodes on the \textsc{Temporary Node Palette}. In this way, AI took on the work — which was shown to be burdensome to users — of continuously noting down the key points in the conversation. Users create links with AI-generated nodes to build a conversational map by themselves. The workflow of \HumanMap is shown in Figure \ref{fig:workflow}(left).

\subsubsection{\AIMap: AI generates the small dialogue maps based on conversation chunks.} \label{AI-Map}
In \AIMap, the AI will first generate the nodes and then generate small dialogue maps after identifying a topic chunk, as detailed in section \ref{segment}. %The mechanisms of identifying topics and generating small dialogue maps are described below.  
Users can freely edit the AI-generated map in any way they want. Given that the synchronicity of the AI-generated content is critical to users, in \AIMap, we first display the AI-generated nodes in the \textsc{Temporary Node Palette} in real-time. At the same time, the system works on creating the dialogue maps in the backend.
% The design consideration of how large the AI-generated map should be and when it should provide a map to users was iterated based on our pilot study.  Users will create and edit the AI-generated small map to make the full map for capturing the conversation. 
The workflow of \AIMap is shown in Figure \ref{fig:workflow} (right).

\textbf{Segment the Conversation into Topic Chunks} \label{segment}
%As mentioned above, the nodes are generated at the turn-taking level, which may not help provide a higher-level conversation overview. To helmeet the needs of users to get the topic-level information of the discussion following the time (D1), we provided a \textit{Topic Timeline} to show the linear information in different granularity (D2). 
%To enable AI to generate a small tree that makes sense semantically, we designed mechanisms to identify topic changes in the conversation. 
%Once the AI detects a topic change, the user will see a small dialogue map for the previous topic that pops up on the canvas. 
The system takes two consecutive turns as the input. It prompts GPT4 to identify if the new turn is a continuation of the previous turn with the same topic or initiation of a new topic. In \AIMap, the topics are also visualized on the \textsc{Topic Timeline Panel}.
% and there will be white space between time blocks to show a new topic identified.

\begin{figure}[htbp]
    \centering
    \includegraphics[width=\textwidth]{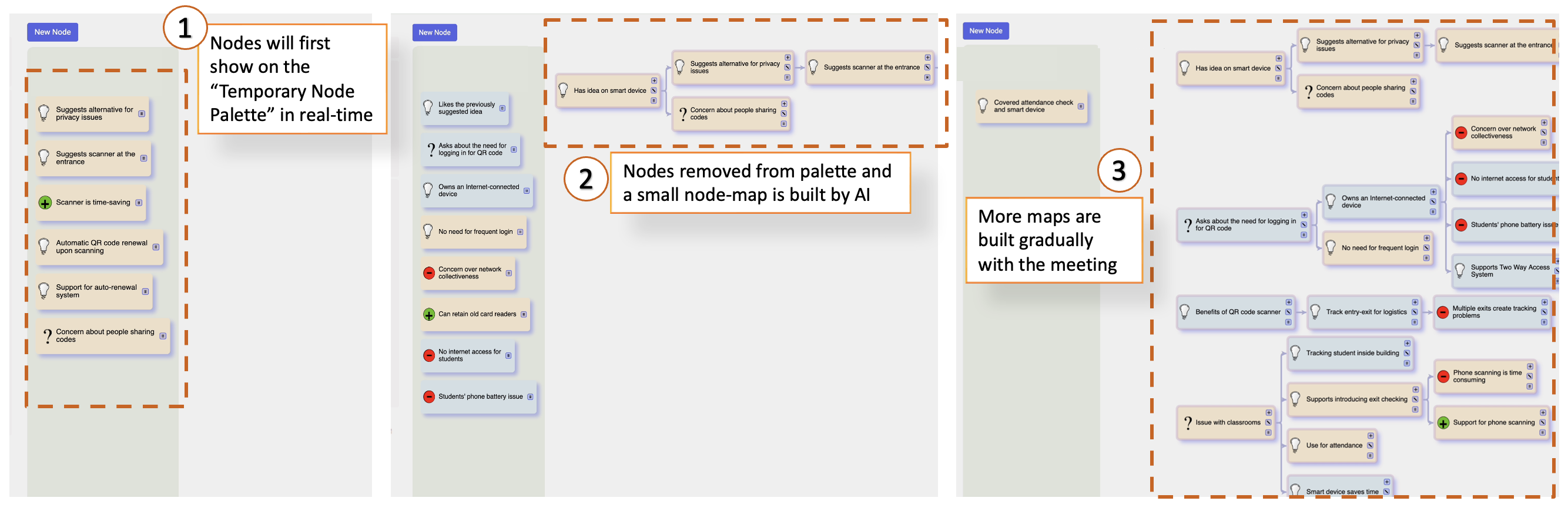}
    \caption{\textbf{\AIMap provides incremental generation of conversation maps so that users perceive the AI-generated content to be in real-time and digestible.} (1) Summary nodes are generated in the \textsc{Temporary Node Palette} in real-time. (2) When a new topic is detected, the nodes in the previous topic are sent to generate a dialogue map. (3) As the conversation progresses,  more small dialogue maps are generated.}
    \label{fig:AI-Map}
\end{figure}

\textbf{Generate small conversation maps using the topic segment}
MeetMap then prompts GPT4 to generate a dialogue map for a given topic segment with the nodes in that topic. The system automatically moves the used nodes from the \textsc{Temporary Node Palette} and presents the nodes with links on the \textsc{Canvas View}, shown in Figure \ref{fig:AI-Map}. 
% 
% (often with several turns).  
% To present the information incrementally. The AI-generated nodes will be first added to the \textsc{Temporary Node Palette} in real-time to enable users to process the node information contextually and in a timely manner. Then, after there is a new topic identified, we prompt GPT4 to generate maps for the nodes identified in the previous segment and then automatically move the nodes from the \textsc{Temporary Node Palette} and present the nodes with links on the \textsc{Canvas View}, shown in Figure \ref{fig:\AIMap}. 
% Users are provided with the same suite of interaction features (as shown in 4.2.1) in \AIMap as \HumanMap to flexibly edit and refine the AI-generated map. 

.

 %Another challenge users mentioned was that a dialogue map fully generated by AI may not facilitate active engagement. To address this challenge, we designed two variants of the system, \HumanMap and \AIMap, to further explore how people would react to different levels of AI assistance \textbf{(D4)}. The design of \HumanMap incorporates users' feedback on how they want to be more active in the process of creating dialogue maps. 

\subsection{Implementation}
For the front-end of the MeetMap, Zoom Web SDK \footnote{https://developers.zoom.us/docs/meeting-sdk/web/} was used to support videoconferencing (as shown in the Figure \ref{fig:overview}, (1) \textsc{Video Panel}), and go.Js \footnote{https://gojs.net/latest/index.html} was used to build the \textsc{Map Creation Panel}.
We used Django Channels and Django-Redis to handle the real-time updates in MeetMap. 
% If a user is muted in the video panel, MeetScript will automatically turn off the audio recording and close the real-time transcript for that user. 
We used the Microsoft Azure SpeechSDK \footnote{https://azure.microsoft.com/en-us/services/cognitive- services/speech-to-text/} to provide transcription. %(the recognition accuracy is more than 90\% as reported) \cite{xu2021benchmarking}. 
MeetMap collects user audio from the client-side browser and then sends the transcription result to the database. Once turn-taking happens, MeetMap calls the GPT4 API to generate nodes as well as identify the topic in both MeetMap variants. Once a new topic is identified, the GPT4 API will be called again to generate links in the AI-Map condition. More details of the pipeline and prompts for generating the map are shown in the Appendix \ref{appendix:prompts}.  
%  D1. intuitive representation of discussion structure following the time manner. 

\section{Evaluation Study}
% \subsection{Participant Recruitment}

% - within-group 
% Compare Zoom + transcript with dialogue mapping

% Compare different levels of AI in dialogue mapping

% task description
%     break
%     decision+brain storming

% survey:
% - usefulness
% - task load

% interview: 
% - how? when?
% - comparison

To understand how MeetMap supports people keeping up with and making sense of the conversation, we performed an IRB-approved evaluation study with three system setups, including the two MeetMap variants and a business-as-usual \Baseline setup. We aim to answer the following research questions:
\begin{enumerate}
    \item RQ1: How AI-assisted collaborative dialogue mapping influenced meeting experiences compared to business-as-usual meeting setups?%When meeting attendees are tasked to collaboratively create dialogue maps during video meetings, how would that affect their meeting experiences compared to business-as-usual meeting setups?
    \item RQ2: How did the different levels of AI assistance influence users’ interaction behaviors and attitudes towards creating dialogue maps during meetings? %When using AI to facilitate a real-time collaborative sense-making task, i.e., creating dialogue maps to understand the conversation, how did the different levels of AI assistance influence their interaction behaviors and attitudes towards creating dialogue maps during meetings?
    %     \begin{enumerate}
    %     \item How do users like the different levels of AI assistance they received during their discussion process? Whether the different levels of AI assistance influenced their interaction behaviors and attitudes towards creating dialogue maps during meetings? 
    %     \item How do users collaborate with each other on the dialogue map with AI scaffolds?
    %     \item How do users perceive the different AI assistance levels and use the AI-generated content in human-human communication?`
    % \end{enumerate}
\end{enumerate}
\subsection{Study procedure}

\subsubsection{Participant Recruitment}
We recruited 20 participants from the University of Michigan mailing lists and divided them into $10$ groups, with $2$ participants in each session. The participants' demographic information is shown in the Appendix \ref{demographic}. The selection of dyadic meetings can help assess how the scaffold in MeetMap can help people in meetings that require rapid information exchange and high cognitive demands. 
All participants were considered to have no hearing or reading difficulties. 
Each study session lasted about 150 minutes, and participants were compensated with an hourly rate of \$$15$. 

\subsubsection{Discussion Tasks}
To encourage information exchange and collaboration, we used the jigsaw method to design discussion tasks \cite{aronson1978jigsaw}. Each participant is presented with a unique point of view, and they have to discuss with each other to reach a consensus on their decisions. To avoid additional difficulty in understanding the topic, we designed $3$ tasks related to school life, including reevaluating attendance checking in university classes, installing smart devices for university buildings, and enhancing mental health services on campus. We designed $2$ discussion agendas for each task to make the discussion concrete and encourage participation. The order of the tasks was counterbalanced for each condition. The tasks used in this study were attached in Appendix \ref{task}.
% These meetings usually last for hours with multiple agendas, and there will be breaks between agendas to organize the discussion \cite{Microsoft}. Thus, having $2$ agendas in the study helps us understand how users use MeetMap to scaffold their discussion both within and across different agendas.  

% , the number of discussions on each task was mostly the same.

\subsubsection{\Baseline Condition}
% Assuming that AI map and \HumanMap were introduced in system design

We designed a \Baseline condition that resembled business-as-usual meeting scenarios. %We enabled meeting transcripts and summaries in the baseline condition to facilitate note-taking and meeting understanding, which were commonly adopted in both educational and industrial meetings \footnote{https://www.zoom.com/en/blog/ai-note-taker-save-time-stay-focused/}. Specifically, 
We connected Zoom with Otter.ai \footnote{https://otter.ai}, a popular Zoom add-on to provide real-time transcripts and automatic summaries for meetings. % which were commonly adopted in both educational and industrial meetings \footnote{https://www.zoom.com/en/blog/ai-note-taker-save-time-stay-focused/}. %as shown in Figure \ref{fig:otterai}.
The transcripts are sent for a summary generation and presented in the summary panel in Otter.ai \footnote{https://help.otter.ai/hc/en-us/articles/5093383818263-Automated-Live-Summary-Overview}. AI presented the summaries as chapters, with several key points and a title for each chapter. The chapter is expandable to present a meeting summary with a two-level hierarchy. 
After seeing the summary, the user could click on the summary to navigate back to the transcript. To the best of our knowledge, this was the most commonly used service that can provide summaries during meetings, rather than only offering meeting notes as post-meeting reviews. Besides, the hierarchy of the summary and the capability of tracking back the original transcript help digest information.  The interval for showing a new summary is around 3 minutes, as set by Otter.ai, which strikes a balance between providing more structured and synthesized information and showing information synchronously. %By comparing this with the turn-based summary generation in both MeetMap variants, our study can provide insights into how different methods of information representation of the conversation - varying in both synchronicity and visualization—impact the participants' meeting experiences.
\begin{figure}[htbp]
    \centering
    \includegraphics[width=\textwidth]{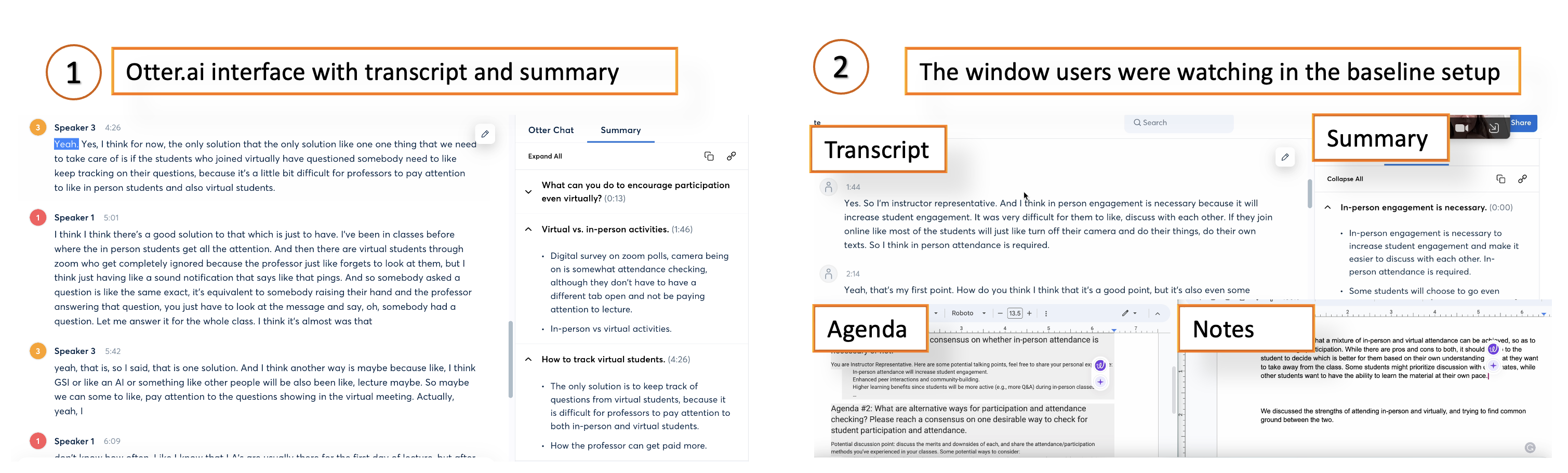}
    \caption{\textbf{The \Baseline condition.}  (1) \textbf{\textit{Otter.ai}} shows real-time transcript (left), and the key points (right) are summarized by AI.  (2) In the baseline condition, three windows are opened and arranged for the participants. a) The otter.ai window is on top to show the real-time transcript and AI-generated summaries. b) The task agenda is placed at the bottom left. c) The shared document is on the bottom right.}
    \label{fig:otterai}
\end{figure}

As shown in Figure \ref{fig:otterai}, the windows of the task and each tool were arranged so that all of them could be read simultaneously. This avoids additional load on the participants when looking up and switching windows during the discussion. % With this setup, Users can view the discussion task and video meeting window while collaboratively taking notes in a shared Google Doc with editable AI-generated transcripts and summaries. They can easily refer to, copy, and paste this content as needed, similar to \HumanMap, where users decide whether to use AI-generated nodes. This setup offers comparable external support for creating shared representations as MeetMap.
The \Baseline condition shows how AI was introduced in daily group meetings in a business-as-usual setting, with an editable and retrievable live transcript open and a bullet point meeting summary enabled and presented linearly. We argue the \Baseline is comparable to the two MeetMap conditions because all provide AI assistance for creating shared representations with dedicated designs to help make sense of the AI-provided information: MeetMap uses AI-generated nodes, while Otter.ai offers structured summaries, both facilitating information digestion through different levels of granularity and navigation mechanisms. By comparing this \Baseline setting to MeetMap, we can see if the AI-assisted dialogue mapping can aid in sense-making during group meetings. %Furthermore, both the baseline and the two MeetMap variants give users similar control over editing, modifying, and using AI-generated content. The difference is in the timing of the introduction of AI-generated summaries and the effort required by users to generate a shared note that can aid collaborative sense-making.  Comparing the three can help us answer the question of when and how AI should be introduced to highly synchronous human-human communication so that it actively engages people in cognitive activities while preserving user agency. 

% \begin{figure}[htbp]
%     \centering
%     \includegraphics[width=\textwidth]{baseline.png}
%     \caption{
%     \label{fig:baseline}
% \end{figure}

\begin{figure}[htbp]
    \centering
    \includegraphics[width=\textwidth]{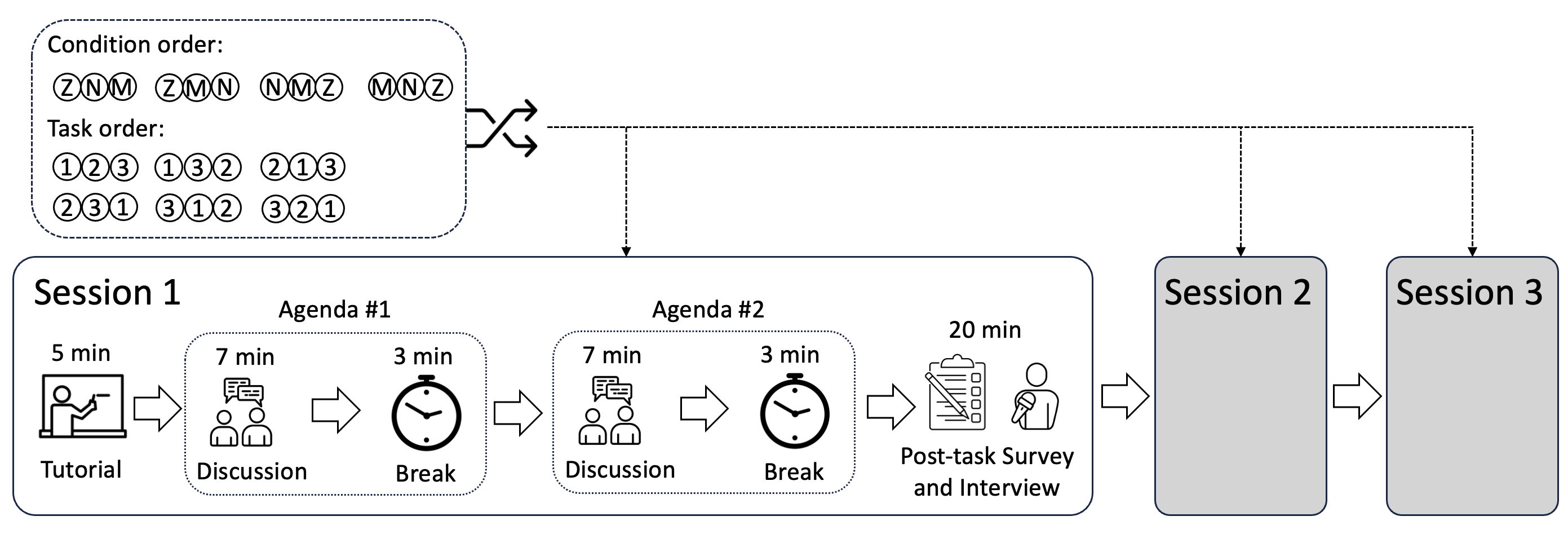}
    \caption{\textbf{Study procedure:} All participants went through the three conditions (M: \AIMap, N: \HumanMap, Z: \Baseline) and the three tasks. The order of the tasks was counterbalanced. The two MeepMap conditions were always done sequentially, but the order between the two MeepMap conditions and the baseline condition were also counterbalanced. A tutorial is given before the task. During the task, the participants discussed two agendas, with a break after each agenda. After each session, all participants answered a post-task survey and had an interview.}
    \label{fig:procedure}
\end{figure}

%Two participants came in person to participate in the study in two separate rooms. This enables us to configure the system to ensure a consistent user experience.
In each session, two participants had a meeting online. Before the session began, we got the participants' verbal consent to record the session. Each participant went through all three conditions. The order of the tasks and conditions was counterbalanced, as shown in Figure \ref{fig:procedure}. %. The detailed task and condition orders are shown }. Participants always do the two MeetMap conditions consecutively so that users can compare the two conditions more directly. 
\subsubsection{Experimental procedure}

%As shown in Figure \ref{fig:procedure}, 
In each task, the participants discussed each agenda for $7$ minutes, followed by a 3-minute break for a recap. They are encouraged to take notes or interact with the map during the discussion. The breaks are introduced since previous studies suggested breaks can help people quickly reflect on the conversation and take time to organize their thoughts\cite{Microsoft}. %, and suggest that taking breaks can improve the group brainstorming quality \cite{schutmaat2023take}. %The rationale for introducing a break in the study is to encourage more interaction with the system to observe the participants' note-taking or map-creation behaviors with AI assistance. %We analyzed users' note-taking and map-creation behaviors during the discussion and the break separately. 

After each task, the participants answered a 5-Likert scale survey about the usability and usefulness of the system. After each task, participants are also asked to complete the NASA-TLX test, responding to six questions on an unlabeled 21-point scale \cite{cao2009nasa}. 
Follow-up questions were asked after the survey about their experiences through a 10-minute semi-structured interview after each task. 
We specifically asked how they perceived and used the AI-generated content during the discussion and during the break. We also asked whether they were satisfied with the AI-generated content, the resulting dialogue maps and notes, and the challenges they experienced. Survey and semi-structured interview questions are provided in Appendix \ref{survey} and Appendix \ref{InterviewQuestion}, respectively.

\subsection{Data Analysis Methods}
\subsubsection{Log data analysis} We analyzed the note-creation and note-checking behaviors between the two MeetMap and the baseline conditions. 

The note-creation behaviors included adding, editing, and deleting nodes and maps in the two MeetMap conditions and creating, editing, and deleting notes in Google Docs in the \Baseline. For the two MeetMap conditions, the MeetMap system logs user interactions with the nodes and the map. 
For the \Baseline, the edit history of the shared document was used to understand user behavior. Two authors watched the recordings to log what changes each participant made to the shared document. They treated each bullet point added by users as new notes and counted the number of times users deleted or edited notes on Google Docs.
Specifically, if a user took notes using short phrases, each short phrase that conveyed a different discussion point was counted as one note-creation activity. If a user wrote a sentence summarizing multiple ideas and reasons, researchers counted each idea and each reason as a note-creation behavior. Additionally, other note-creation behaviors, such as changing the order of notes or editing existing notes, were tracked. 

The note-checking behaviors counted users' interactions to check and read the content. This included locating a node on the map by clicking on the timeline, panning across the map using the mini-map, scrolling the node palette to read more information, and double-clicking the node to check the transcript in the two MeetMap conditions. In the \Baseline, note-checking behaviors included users scrolling on the transcript or their collaborative notes, clicking to expand the summary on otter.ai, and clicking the summary to jump back to the original transcript.

Although users interacted differently in MeetMap and the \Baseline, the note-creation and node-checking behaviors were measured using a comparable granularity for consistent computation.

% \subsubsection{Discussion Analysis} 
% \subsubsection{Measurement of understanding} To establish a reference point, one author watched the discussion recordings and generated a list of bullet points as the "ground truth", encompassing the final decisions and their main reasonings reach this conclusion. Then two authors independently coded all participant responses and assigned scores to each response.
\subsubsection{Survey data analysis} %To address RQ1, we compare the difficulty level of each condition using NASA-TLX and the satisfactory level by asking them to rate the quality of the final results and the probablity to use the system in the future. 
The survey data was evaluated using the Friedman test because the data did not meet normality assumptions. For each question, we established a null hypothesis that no statistically significant difference exists among the three conditions. We conducted the Wilcox signed-rank post-hoc test to compare the differences between the two conditions.

\subsubsection{Interview analysis} The interviews were transcribed, and two researchers used the Affinity Diagram  \cite{ lucero2015using} to analyze the data. In the analysis, two researchers rearranged all quotes iteratively based on emerging affinity to one another through communication and critique. We grouped users' feedback, including how they create notes/dialogue maps with the help of AI both in-situ and post-meeting, why or why not the scaffold of MeetMap is helpful, their preferences, and concerns about incorporating AI in synchronous meetings.

\subsubsection{Video data analysis}  %In addition to the log, survey, and interview data, 
We qualitatively analyzed the video recordings to observe participants' collaboration behavior and their interactions with the system. The video analysis was used to understand the nuances of interactions and behavior patterns that might not be captured through log data or self-reported measures.
Two researchers independently watched the video recordings and wrote memos about how people took notes during the discussion and the break, how they collaborated with each other, and how they used and edited the AI-generated content. After that, they came together to discuss the memo and used it to complement and explain some findings from log analysis and interview analysis.

% semantic analysis + affinity diagram
% log data
% turn taking / understanding check / xxx
% cognitive load (questionare) NASA

\section{Findings}
The comparison between the two MeetMap variants with the \Baseline demonstrates the usability and usefulness of MeetMap in helping people keep track of and understand conversations in real-time (RQ1). The participants appreciated the flexibility of MeetMap in allowing them to structure the conversation visually. % Of the three conditions, users had more cognitive bandwidth in \HumanMap to take notes during the discussion due to the synchronous and concise display of the AI-generated content, which enabled them to read and act upon it quickly. 
The comparison between Human-Map and AI-Map showed that users in Human-Map were more motivated and perceived a higher level of control and agency to create and actively modify the dialogue maps (RQ2). Furthermore, users reported a lower tolerance for AI errors when they felt they put more effort into creating the map.

\subsection{How AI-assisted Collaborative Dialogue Mapping Influence Meeting Experiences in Comparison to Business-as-usual Meeting Setups}

We first present results on how MeetMap (Human-Map and AI-Map) helps people make sense of the discussion compared to the \Baseline condition.  

\subsubsection{MeetMap helps individuals keep up with and make sense of the meeting in real-time }
\label{understanding}
\begin{figure}[htbp]
    \centering
    \includegraphics[width=0.95\textwidth]{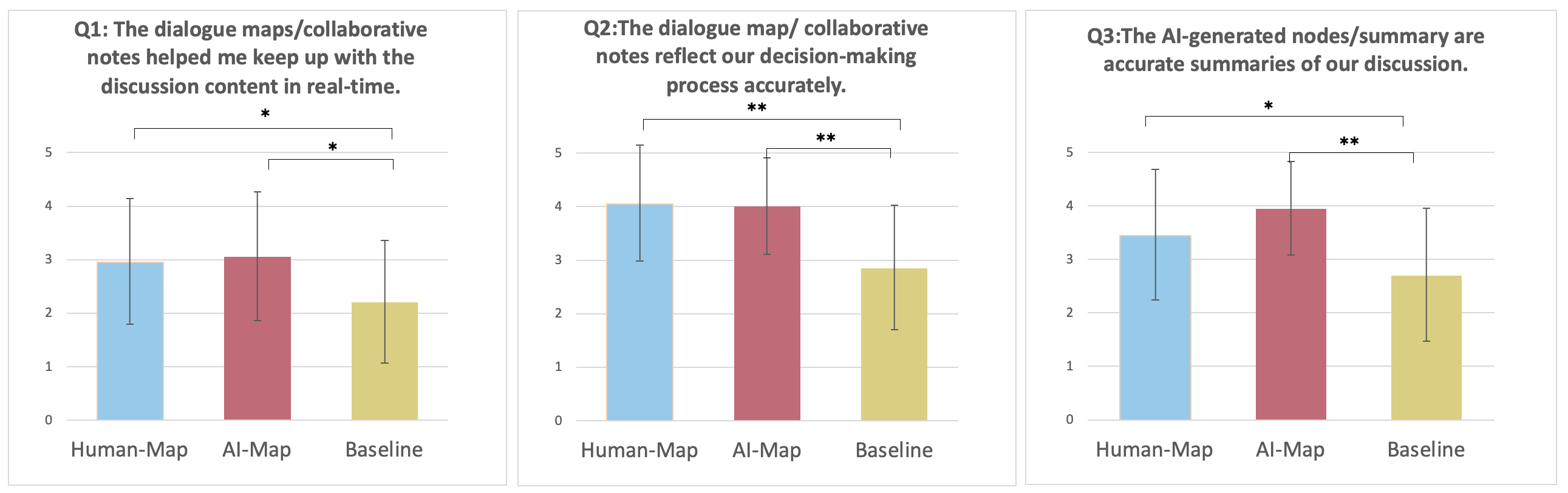}
    \caption{\textbf{Survey questions} 1) Q1: The dialogue maps/collaborative notes helped me keep up with the discussion content in real-time. 2) Q2: The map/the AI transcript and summary reflect our decision-making process accurately. 3) Q3: The AI-generated nodes/summaries are accurate summaries of our discussion. ($*: p<0.05$; $**: p<0.01$; $***: p<0.001$)} 
    \label{fig:survey_Q2_34}
\end{figure}

Users reported that both MeetMap variants helped them keep up with the discussion content better than the \Baseline (p = 0.017 < 0.05, p = 0.46 < 0.05). (Figure \ref{fig:survey_Q2_34}:Q1).
Users thought the dialogue map reflected their decision-making process more accurately in both Human-Map (p = 0.001 < 0.05) and AI-Map (p = 0.003 < 0.05) conditions compared to the shared notes in the \Baseline condition (Figure \ref{fig:survey_Q2_34}:Q2). 
Additionally, users considered the AI-generated summary nodes and dialogue maps to be accurate representations of the conversation in both Human-Map and AI-Map, in comparison to the AI-generated summaries in the \Baseline (p=0.03 < 0.05, p=0.002 < 0.05 (Figure \ref{fig:survey_Q2_34}:Q3).

Qualitative insights revealed that the perceptions were due to the way MeetMap represents linear conversations in a structured, visual, and intuitive manner. %
All users agreed that the visual representation and notation schema facilitated a quick grasp of the content. P2 noted, \textit{``It listed some icons with visual categories to show our opinion and helped me quickly get the main idea and its sub-nodes.''}  
Most users (16/20) ) found the design of MeetMap intuitive for connecting ideas over time. %, particularly mentioning the benefits of being able to "navigate to an idea on the map through the timeline" and "organize ideas semantically on the map."
As P2 stated, \textit{``We could connect information from the second agenda into the previous agenda (P2 pointed to the agenda topics on the timeline), and synthesize ideas (P2 pointed to the arrows on the Map Canvas)."} 
Additionally, the right amount of extra information was crucial for in-situ understanding. All users (20 out of 20) believed that a complete transcript provided excessive information in the baseline condition; %P5 commented, \textit{"What we said is too tedious to put on the note."} %Additionally, four users noted that the transcript's accuracy reduces usability in the \Baseline, as exemplified by P10: \textit{"Here are some grammar issues in my raw speech, and also with many connection words. I don't want to put them in my notes."} 
In contrast, all users (20 out of 20) preferred the concise content in the AI-generated nodes in MeetMap. P9 observed, \textit{"The nodes took out every point. But every node is really short, so I can quickly glance at it."
}

\subsubsection{MeetMap fosters team consensus building and facilitates subsequent discussion}
\begin{figure}[htbp]
    \centering     
    \includegraphics[width=0.33\textwidth]{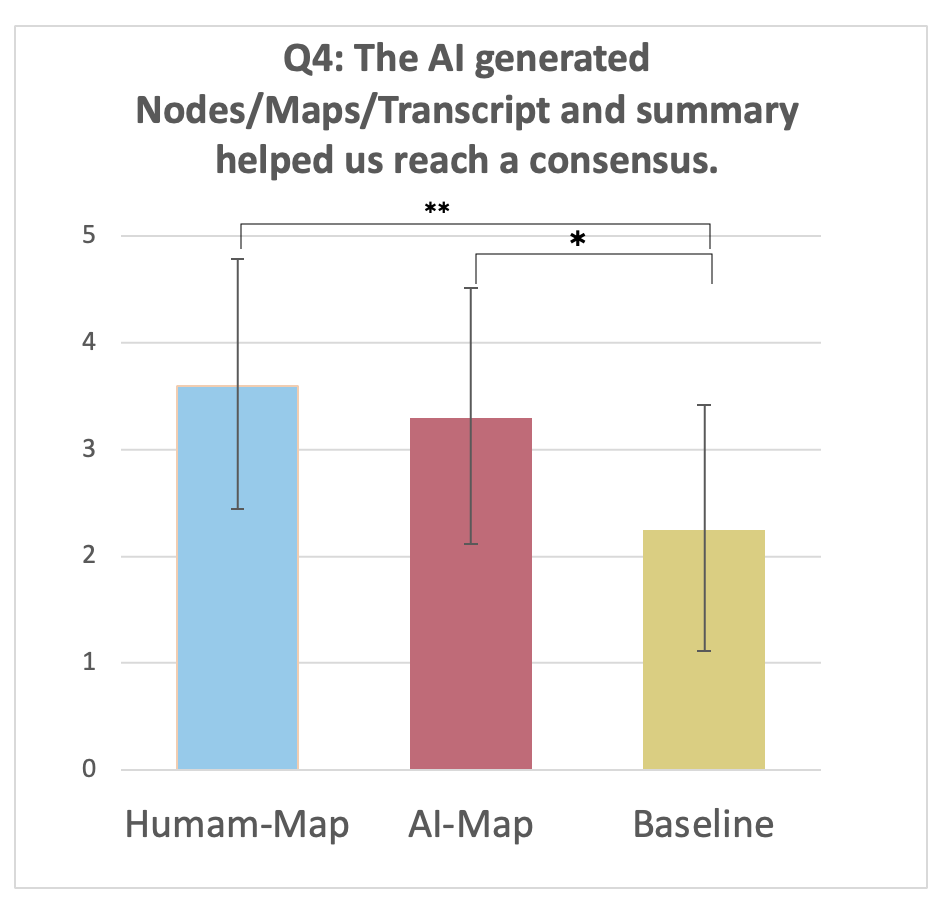}
    \caption{\textbf{Q4: The AI-generated Nodes/Maps/transcript and summary helped us reach a consensus. } Human-Map and AI-Map show significantly higher ratings in helping people reach a consensus. The error bars represent standard deviations. ($*: p<0.05$; $**: p<0.01$; $***: p<0.001$)}
    \label{fig:survey_Q2_26}
\end{figure}

%When assessing the role of dialogue maps and the summary in the baseline condition in achieving team consensus .  
Users rated Human-Map (p = 0.003 < 0.05)  and AI-Map (p = 0.004 < 0.05) significantly higher in supporting teams to achieve consensus compared with the \Baseline, as shown in Figure \ref{fig:survey_Q2_26}:Q4). 

Some users (8/20) indicated that MeetMap helped build shared understanding and facilitated more structured subsequent discussions among teams. The AI-generated nodes in MeetMap were perceived as objective mediators to address misunderstandings. 
For example, during session three, P6 noticed a node was mistakenly categorized as a "Con" instead of a "Pro" and said, \textit{``Oh, I thought this should be a benefit of this idea rather than a negative one. Maybe we have some misunderstanding, and do we want to discuss it further?"} 
Users (P7, P10, P11) found it less confrontational to address misunderstandings using AI-generated nodes,  \textit{``I didn't feel there was any way to clear up a misunderstanding (in \Baseline). I could say that directly, but that felt more combative than correcting a misunderstanding in the other one (in MeetMap)."}. 
In addition to solving the misunderstanding, there were 3 out of 10 groups of participants who actively created nodes to guide the conversation in Human-Map;
as P2 said, \textit{``I would try to create two top-level nodes, which helped me organize our discussion based on these ideas."}.
Some users (7/20) believed that the notation schema in MeetMap helped them reflect and promoted alternative perspectives. As P17 said, \textit{``In MeetMap, if I found one idea with three pros and one con, I would ask the team to consider whether there are more cons for this idea.''} 
                                                                                                            
\subsubsection{MeetMap users have more bandwidth to create and read their shared notes during the conversation.} \label{cognitive}

\begin{figure}[htbp]
        \centering
        \includegraphics[width=0.5\textwidth]{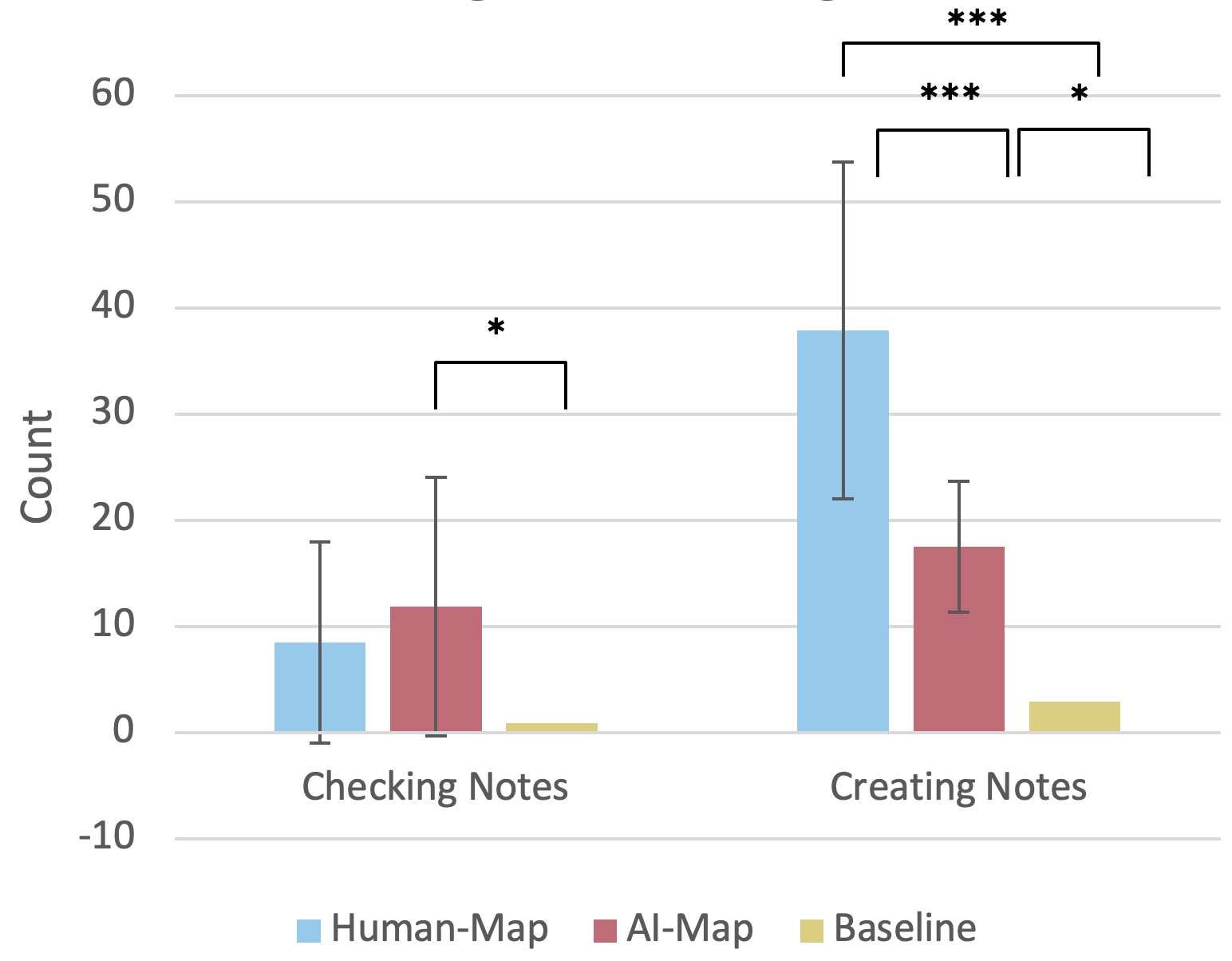}
        \caption{\textbf{Users created more notes with MeetMap. }Users result in a significant increase in note creation and viewing in MeetMap. The error bars represent standard deviations. ($*: p<0.05$; $**: p<0.01$; $***: p<0.001$)}
    \label{fig:usage}
\end{figure}

We found that Human-Map users had significantly more note-creation behaviors than AI-Map ($p=0.0001<0.001$), who also had more note-creation behaviors than the \Baseline ($p=0.015<0.05$), as shown in Figure \ref{fig:usage}. Besides, users had significantly more interactions to navigate and check the note content in AI-Map than that in the \Baseline ($p=0.035<0.05$). 

Creating more notes during the conversation may possibly incur a higher cognitive load on the users.
% One possible trade-off for having more note creation and other interaction behaviors is that it may require higher cognitive demands on the users. 
However, based on our NASA-TLX survey, participants did not report a higher task load when comparing two MeetMap conditions and the \Baseline, as shown in Appendix \ref{cognitive-load-survey}. Despite the similar level of cognitive load as reported, users interacted more with the map and created more collaborative notes in the two MeetMap conditions. This increased interaction likely enhanced users' real-time understanding of the conversation, as reported in Section\ref{understanding}. %The bandwidth to interact frequently with the map, navigate and check content, and create notes provided users with a dynamic and engaging way to process information as it was being discussed. 
% This conclusion is echoed in the qualitative findings. 

%Some users (6/20) in the baseline condition reported that they would miss what others were discussing when taking notes. As P8 stated in an interview, \textit{"I was trying to finish this sentence. And I didn't realize it was my turn until he stopped speaking.
%"}  %Furthermore, one challenges they faced during note-taking is  ``the difficulty to predict or confirm the focus of a conversation to be noted'' (P3, P8, P12, P20).
%Users believed that the AI-generated nodes in MeetMap solved the difficulty.
%They found the direct manipulation of organizing linear information with visual structure was much easier than creating content from scratch.
%As P7 said, \textit{``You just see all the nodes, it's like block-based programming, then you just have to arrange them, or link them or maybe add a few things. That's much simpler than remembering the whole conversation and taking notes on your own".}

Most users (18/20) thought the synchronicity of the immediate summary nodes made taking notes in real-time possible in both MeetMap conditions,
like P10 said, \textit{"If the nodes are created at the same time that we're having the conversation, my mind is still on the topic being discussed, so it was easier for me to use it."} 
7 users especially appreciated the way that AI first generates the nodes and then combines them as maps in AI-Map, which reduced their uncertainty about the AI ability, as P8 said, \textit{"I know it is working once I see the nodes pop out after I speak, then it generates links, So I don't need to worry about missing anything. "} 
On the other hand, most users (14/20) were confused by the long and unpredictable waiting time of the AI summary in the baseline condition. %; a%lthough Otter.ai generates the summary around 3 minutes, it's too long for users to perceive AI is working and predict a time the summary will be generated.

\subsubsection{The pair of participants have more balanced contributions to the dialogue maps in MeetMap than Zoom.}
\begin{figure}[htbp]
    \centering
    \includegraphics[width=\textwidth]{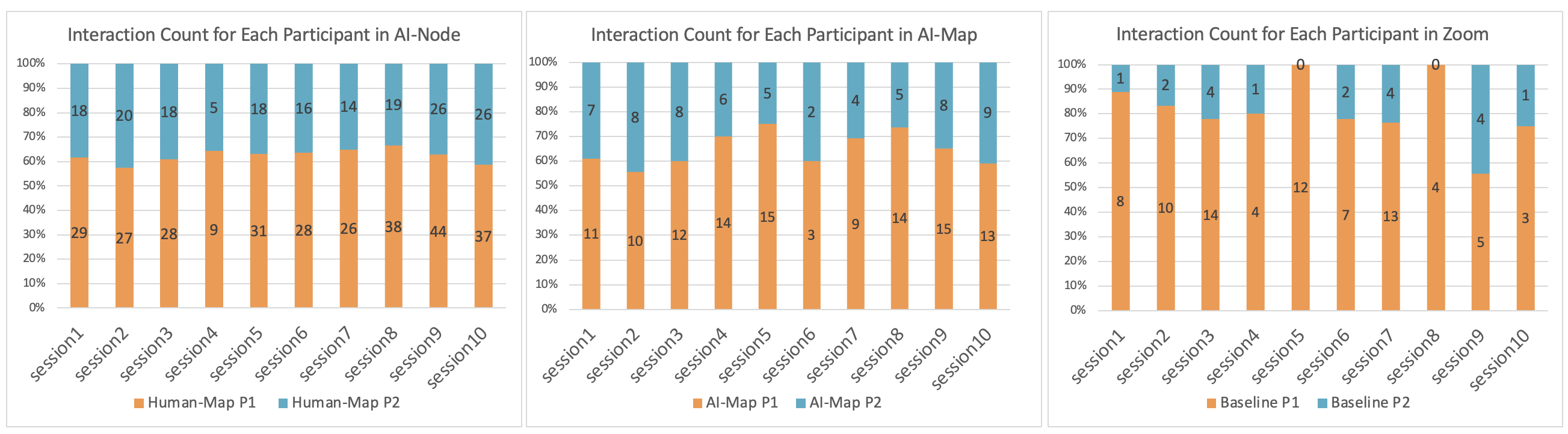}
    \caption{\textbf{We analyzed the relative number of note-creation behaviors between the two team members in each group.} The note-creation behaviors included adding, editing, and deleting nodes and maps in the two
MeetMap conditions, and creating, editing, deleting notes in Google Docs in the Baseline.  Users in the Human-Map and AI-Map conditions show more balanced collaborative note creation behaviors in comparison to the \Baseline.}
    \label{collaboration}
\end{figure}

We analyzed how each member in the pair interacted with the maps and notes in all conditions.
%In each group, we presented the number of notes taken by two participants and the proportion between them. 
As shown in Figure \ref{collaboration}, in the two MeetMap conditions, both participants in each group actively took notes, showing more balanced collaborative note-taking behaviors. In contrast, in the Zoom condition, the contribution between the two members in each group was much less balanced.

% each group tended to have one person dominate the collaborative note-taking.

%We summarized the collaborative patterns from the video analysis. In both MeetMap settings, users actively participated in note-taking and spontaneously adopted different roles within the group when dialogue mapping. In 5 out of 10 groups, one member adjusted the AI-generated content while the other added overarching themes to create a comprehensive structure. All groups attempted to merge their notes into a comprehensive map to reflect the meeting's overall trajectory. %This shows that the design of MeetMap facilitated spontaneous role assignment.
%During the \Baseline sessions, we observed a different dynamic. Some groups (4/10) had one participant responsible for taking the discussion notes, while the other person only watched and provided verbal suggestions. Other groups (6/10) had both members taking notes, but they typically wrote in separate paragraphs and did not try to combine their content.

Video analysis of the meetings also showed that users in the MeetMap conditions demonstrated more collaborative and co-creation behaviors, e.g., in all groups in the MeetMap conditions, the two team members merged their notes and collaboratively edited the dialogue maps.
In contrast, in the baseline condition, participants were more likely to take individual notes and did not integrate the notes. 
For example, in 4 out of 10 groups in the baseline condition, one participant took notes while the other watched and provided verbal input. Both members took notes independently in the remaining 6 groups, typically writing in separate sections without combining their content.

In the post-interviews, we probed the reasons behind the observed collaborative behaviors. 
First, users found that indicating the speaker on the dialogue map in different colors helped users feel more ownership of their contributions. 
% visualization to indicate speakers helped 
Users (8/20) shared that they were more likely to engage with nodes that reflected their own contributions, as indicated by different colors in the dialogue maps. 
% This feature helped users feel ownership over the notes, encouraging active participation.  
P20 said, \textit{"So I can see those ideas were said by myself, so I am the one to organize those ideas."}  
However, in the baseline condition, users had less clear expectations around who to take notes and what they should note, as P6 said, \textit{"I was not sure if she would write something "} 
Besides, users found it easier to refine AI-generated content rather than altering notes created by their peers. 
%This fostered a more collaborative environment, as modifying AI-generated nodes felt less intrusive; 
As P1 said, \textit{"I didn't feel like I was changing other people's work while working on the AI-generated nodes. "}  
 AI-generated nodes were perceived as a neutral foundation for establishing common ground. As P13 explained: \textit{"It’s like we’re all shaping a common ground, using AI's input as blocks and then working toward an integral dialogue map together. It is less personal and more about improving the overall result."}

While we observed co-creation behaviors on the dialogue maps in both MeetMap conditions, most groups (7/10) collaborated on the map without verbal coordination. This sometimes led to unexpected changes in node content or map structure made by others. For instance, P12 noted, \textit{“It got a bit chaotic when we all started dragging and organizing nodes without a clear strategy.”}
%participants still encountered some collaboration challenges. 
%some people still didn't know how to coordinate with each other to generate a dialogue map without explicit guidance. %Although we found participants in some groups would discuss the potential structure they would use in the map, other 
% \subsection{What motivates or hinders users from collaborating with AI to generate meeting notes during the discussion}

%We analyzed users' note-taking behaviors during the discussion and the break separately.%A separate analysis would allow us to understand the specific needs users have in taking notes during different stages of the discussion. %infer the cognitive load placed on the users by the system. %
%An assumption is that users can only take notes during the discussion when the system provides the right amount of information that aligns with the user's mental model.  %On the other hand, being able to take notes during the break places a lower requirement on the system design. 

\subsection{How the Different Levels of AI Assistance Influenced Users' Interaction Behaviors and Attitudes Towards Creating Dialogue Maps During Meetings}
In this section, we will examine user experiences in two variants of MeetMap,  \HumanMap and \AIMap. %These variants have different levels of AI assistance. 
We aim to understand how much AI assistance is desirable and useful for people working together to create dialogue maps during discussions. 

\begin{figure}[htbp]
    \centering
    \includegraphics[width=\textwidth]{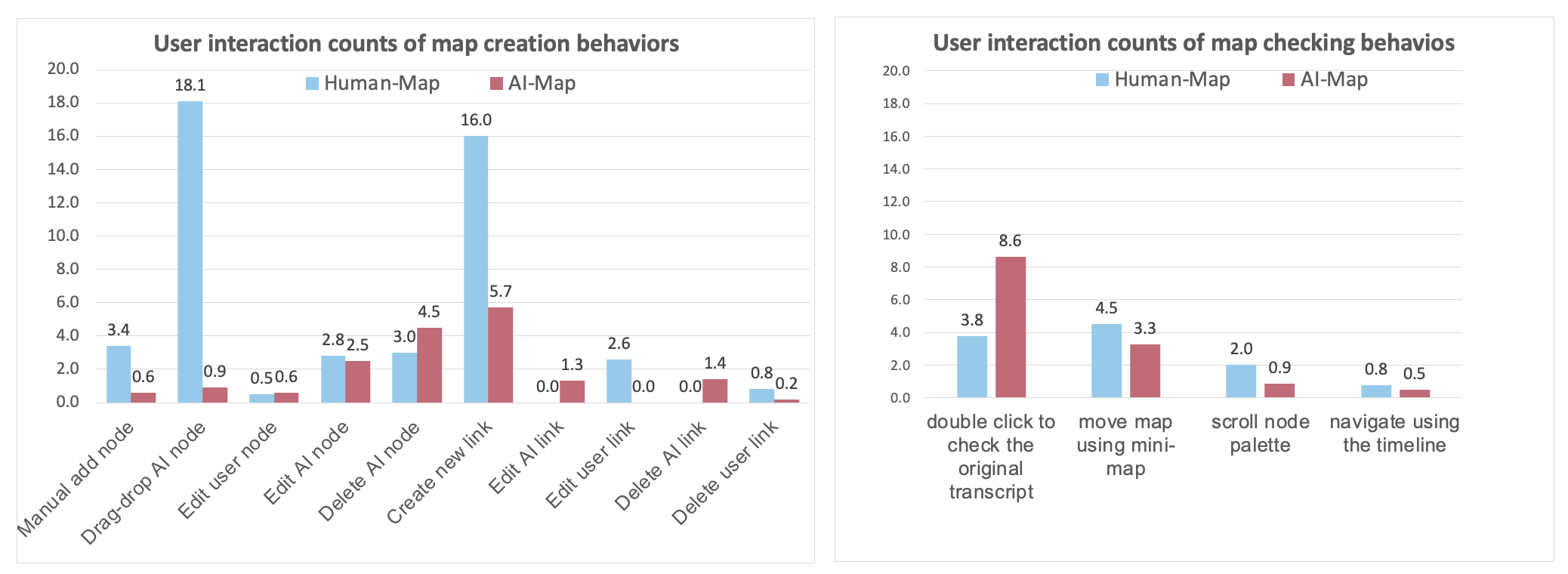}
    \caption{\textbf{The detailed user interaction behaviors in \HumanMap and \AIMap. }  The average count of dialogue map creation behaviors per person \textbf{(left)} shows that users in \HumanMap demonstrated more map creation behaviors, such as manually adding nodes, dragging and dropping AI nodes, and creating new links. \AIMap users showed fewer map creation behaviors but more review behaviors, such as editing the content and the categories of AI-generated nodes, editing the AI-creating links, and deleting nodes or links. The average count of the dialogue map checking behaviors \textbf{(right) }showed that \AIMap users frequently checked the original transcript behind an AI-generated node.}
    \label{comparison}
\end{figure}

Figure \ref{comparison} shows the intensity of user interaction in both MeetMap variants by detailing the specific map creation and review behaviors.  
% detailed counts of the dialogue map creation behaviors of the two MeetMap variants on the left. 
As shown in the left figure, users in \HumanMap demonstrated more map creation behaviors, such as manually adding nodes, dragging and dropping AI nodes, and creating new links. 
On the other hand, users of \AIMap showed fewer map creation behaviors but more review behaviors, such as editing the content and the categories of AI-generated nodes, editing the AI-created links, and deleting nodes or links.
As shown in the right figure, \AIMap users checked the original transcript of the nodes more frequently by double-clicking the nodes.

% \begin{figure}[htbp]
%     \centering
%     \includegraphics[width=.5\textwidth]{viewing.png}
%     \caption{\textbf{The viewing behaviors in \HumanMap and \AIMap.}}
%     \label{viewing}
% \end{figure}

\subsubsection{\HumanMap provided more agency and control for users to dialogue map in real-time }
In \HumanMap, most users (13/20) reported having more agency and control to use the AI-generated nodes, as mentioned by P14, \textit{"I saw it was generated here (Temporary Node Palette), and then it was my responsibility to move it to the map to make it useful." }% The synchronicity and the simple interaction of dragging and dropping nodes to the maps allow users to feel more in control, fostering active engagement in shaping the dialogue map with AI. 
In contrast, in \AIMap, users preferred to simply read the AI-created maps during the discussion, leaving potential post-editing until after the discussion. 
This was because there was uncertainty about when the AI-generated maps would appear on the Map Canvas and when and how the AI-generated nodes on the Temporary Node Palette would be used to create the dialogue maps. P12 explained, \textit{"I was not sure which nodes would be added to the map by AI, so I hesitated to add new nodes or modify the AI-generated map during the discussion."} %Such delays and unpredictability in AI assistance can reduce the users' sense of control.

\subsubsection{Users in \HumanMap and \AIMap made sense of the dialogue maps differently.} %\AIMap and \HumanMap exhibited two distinct ways of interacting with AI-generated content, each demanding different cognitive processes from users.
% Our findings indicate that both systems require active cognitive processing but cater to different user preferences for making sense of information. 
% We found that 
6 users (6/20) shared that they needed to put in more cognitive effort in \HumanMap to review the content in each node and think about how to organize them than they did in \AIMap, as P16 said,\textit{ ``I think when the AI tool grouped everything together (\AIMap), it was easier for me just to scan the information and confirm if it was okay and that's it. But in this case (\HumanMap), I had to go through the entire discussion again and connect stuff from beginning to end, so it took me longer to think, compare, and remember."} 
%In \HumanMap, users organized nodes into a structured dialogue map based on AI-generated summaries. This task involves actively synthesizing and structuring information, requiring high-level cognitive processes.
In contrast, 8 users reported that \AIMap simplifies the initial organization of content through AI-generated maps, yet users remain critically involved in refining and verifying AI-generated output. %This task focuses on adjusting inaccuracies and ensuring the relevance and coherence of the information presented by AI.  
P3 said, \textit{``It was more challenging to understand if I didn't create it myself."} 
%Our findings indicate that both systems require active cognitive processing but cater to different user preferences for making sense of information. \HumanMap aligns with users who prefer a constructive approach, involving active building and organization of information. In contrast, \AIMap suits users who excel in analytical tasks, focusing on refining and optimizing pre-structured content.

\subsubsection{Users shared that they had a better understanding of the conversations when they co-created the dialogue maps in \HumanMap} 
Many users (12/20) reported that they preferred intensive cognitive engagement in creating dialogue maps in \HumanMap, despite it requiring more effort. These users believed that organizing the information by themselves was helpful for making sense of the content and led to better decision-making, even though the process could be cognitively demanding. P20 said, \textit{``So it actually takes a lot of cognitive load for us to think about how to organize the nodes. But I think personally it's helpful, and we need to do that. 
} 
Most users (14/20) commented that they better understood the session and could recall the content more easily after creating the dialogue maps in \HumanMap. %, as P1 said, \textit{``The structure makes it  easier for me to recall what has happened." }. 
In comparison,  some users (5/20) mentioned that they put less effort into creating dialogue maps and found recalling the discussion and decision-making process more challenging in \AIMap, as P5 said, \textit{``It was harder for me to remember it since it (AI) adds another logic to explain the conversation."}. %This preference suggests that users value the deep understanding that comes from actively organizing and synthesizing information, compared to passively consuming AI-generated summaries. 

\subsubsection{Users had a lower tolerance for AI mistakes when they considered themselves to own the human-AI collaborative output}
Users (6/20) shared that %when they spent more effort to refine the AI-generated content, they had a lower tolerance for AI mistakes and demanded better outcomes from the joint human-AI efforts.
their involvement in creating the dialogue maps in \HumanMap made them more critical of AI outputs. Because they felt they were responsible for the final deliverable. 
To create dialogue maps with AI-generated nodes, users would first read and comprehend the AI-generated nodes to ensure the content was accurate before using this node in the dialogue map. P13 said, \textit{``If it's wrong (\HumanMap) and if it's not what I'm trying to express, I have to figure out how to use this one, check what it wants to say, make edits on it, or build a new one."}
In contrast, despite encountering similar inaccuracies, most users of \AIMap were more receptive to AI mistakes. Only 2 users pointed out inaccurate content on AI-generated nodes in \AIMap, and others were quite liberal with AI-generated links. P16 said, \textit{``Since I need to be responsible for the quality of this map, and it's not the tool (\HumanMap). And the other one, it was so easy that I needed to scan it, and I don't mind some errors (\AIMap)".} 

% \subsubsection{Users had mixed feedback on the trust on AI in understanding the logic behind their conversation.}
% In both the \AIMap and AI-Node conditions, users commented that AI did a good job creating short summaries of meeting content. 
% However, users had different opinions on whether AI should generate the dialogue maps directly for them. About half of the users (11/20) preferred to build the connections by themselves. They thought that AI could help record the communication process but could not understand the logic behind the conversation. P5 said, \textit{``I trust the AI a lot more to understand what I'm saying more than I trust it to connect things together." }. %These users believed that organizing the information by themselves was helpful for them to make sense of the content and led to better decision-making, even though the process of creating the dialogue maps could be cognitively demanding. P20 said, \textit{``So it actually takes a lot of cognitive load for us to think how to organize the nodes. But I think personally it's helpful and we need to do that and this is what we usually do in design meetings".} 
% %Other users (9/20) preferred for AI to generate the dialogue maps directly and would like to use the dialogue maps as memory aids for post-meeting review and reflection. 
% %P10 said, \textit{``I think most people, including me, will be happier with the AI generating a basic dialogue map and we just need to shift one or two nodes."} 

\subsection{Challenges of using MeetMap to support collaborative sense-making during meetings}

While users appreciated MeetMap for its ability to facilitate conversation and capture meeting content, several issues emerged that warrant further consideration.

% \subsubsection{Collaboration challenges and the need for more explicit guidance}

\subsubsection{Notation schema need to be adaptable}
Some users (3/20) found that the predefined categories of the dialogue mapping notation schema did not always suit the content of their conversations, limiting their ability to represent the discussion accurately. %As noted by P7, P14, and P15, the schema was sometimes too restrictive, creating confusion. 
P15 mentioned, \textit{``It will be good if I can add a type. Or you can provide many templates for users to choose."}  Participants (3/20) need a system that can adapt to the unique dynamics of different conversations, with the ability to introduce custom categories or choose from a broader range of templates. %We should note that using the "Dialogue Mapping" schema is a practice for exploring the potential of using AI to categorize and visualize conversations that can support in-situ sense-making. In this study, we didn't aim to test the AI's generalizability of categorizing conversations using an on-demand notation schema. We will leave this for future work.

\subsubsection{The need for control over AI granularity}
Users also desired more control over the granularity of the AI-generated content.%, emphasizing the importance of being able to personalize content generation according to their specific needs. 
%The ability to dictate what type and amount of content the AI should generate was vital for participants. 
For example, some users complained about the repetition of the nodes due to either back-and-forth discussion or the imperfect performance of AI in chunking the conversation. % and hoped they could instruct AI to combine similar nodes and only present unique information for them. 
As P9 said, \textit{"there were some nodes which had  similar ideas; it will be great if I can tell AI to include just one."} While users liked to see the visuals of the notes, they wanted to have more control over what information to be recorded and what structure to use to organize the notes. As P12 said, \textit{"Different people map things in different ways. It will be cool if we could decide what kind of structure we want to get in this meeting".}

% \subsubsection{Some users did not trust AI in understanding the logic behind their conversation.}
% In both the \AIMap and \HumanMap conditions, users commented that AI did a good job creating short summaries of meeting content and categorizing turns with notation schemas. 
% However, some users (7/20) preferred to build the connections by themselves since they thought that AI could only help record the communication process but could not understand the logic behind the conversation. P5 said, \textit{``I trust the AI a lot more to understand what I'm saying than I trust it to connect things together." }.
% %Other users (9/20) preferred for AI to generate the dialogue maps directly and would like to use the dialogue maps as memory aids for post-meeting review and reflection. 
% %In these situations, they were not motivated to create the dialogue maps by themselves. 
% %P10 said, \textit{``I think most people, including me, will be happier with the AI generating a basic dialogue map and we just need to shift one or two nodes."} 

\section{Discussion}

Our findings highlight how MeetMap enhances real-time collaborative sense-making during video meetings by combining structured visual representations with synchronized AI assistance, offering valuable insights for designing AI systems that balance automation and user agency. In the following discussion, we explore these insights in detail across three key areas: (6.1) Design Implications for AI-assisted Real-time Sense-making, focusing on how structured visuals and staged information aid navigation and reduce cognitive load; (6.2) AI-assisted Collaborative Dialogue Mapping, examining how AI-generated content promotes balanced collaboration and shared understanding; and (6.3) User Agency and AI Assistance, discussing the importance of withholding AI support to preserve user control and engagement.

% To reduce the overload of introducing human-AI collaborative workflow to help sense-making in synchronous meetings, personalized preferences of how to structure and make sense of the information should be considered. While all users in our user studies preferred MeetMap over Zoom+Otter.ai, this doesn't mean all of them like the way that we use the IBIS notation schema to structure the information for them. 
\subsection{Design Implication for AI-assisted Real-time Sense-making during Video Meetings }

First, using visualizations to enhance the structure of AI-generated content can reduce users' cognitive load, especially when they need to process information in real time. In MeetMap, users appreciated features such as color-coding for different speakers and the dialogue map notation schema, which categorized nodes into four categories(\S5.1.1.). 

Second, MeetMap provides transparency into how dialogue maps are created through visualizing the summary nodes following a chronological order in the \textsc{Temporary Node Palette}, which was applauded by the users and helped them understand the dialogue maps. 
This suggests that establishing a clear connection between multiple representations of information, e.g., chronological and semantic elements, could help users navigate and locate information more effectively  \cite{Jiang2023GraphologueEL, zhang2024ladica}.

Third, delays in using AI to synthesize conversations (e.g., generating summaries and dialogue maps) can affect user experience. To address this, presenting intermediate AI outputs can enhance the system's perceived synchrony and transparency. In MeetMap, the introduction of the Temporary Code Palette was widely praised by users, as it displayed summary nodes in real-time, allowing them to gradually understand the content before the full dialogue map was generated.
The synchronicity of MeetMap to provide concise summary nodes increased users' capacity to consume the information in real time and also reduced users' uncertainty and frustration compared to the baseline with longer waiting times. 
% In AI-Map, people receive synchronous information through the \textsc{Temporary Node Palette} and are shown delayed as maps, suggesting that a staged presentation of information could be helpful for users to consume the information in real-time while allowing the machine to produce the result in the back-end.  
% This middle-layer design for displaying AI-generated intermediate results in real-time could be utilized in other contexts, enabling users to view content instantly while AI processes substantial data volumes, thus reducing potential delays.

Lastly, enhancing users' cognitive engagement is particularly beneficial for cognitively demanding tasks, such as collaborative conversations. While Human-Map offers less automation compared to AI-Map, it was praised by users for providing more opportunities to actively engage with and make sense of the content.
% At the same time, our study shows that through inviting users to actively contribute to create and refine the dialogue maps, it reduces their cognitive load in processing AI-generated content. 

% our study shows that both Human-Map and AI-Map manage cognitive load effectively by encouraging more active interaction in creating shared notes for sense-making  (\S5.1.3). 

% Beyond the visual representation, supporting in-situ understanding in a highly collaborative and synchronous environment also requires careful consideration of how much cognitive load the extra information may place on users \cite{chen2023MeetScript}.

We also identified limitations in the MeetMap system. %For 
The users expressed a desire for a personalized notation schema and diverse visual structures to create the dialogue map (\S5.3.1).  %To address these issues, future improvements could include letting users customize the schema tags to increase flexibility. 
Recent work showed the benefits of LLM-generated templates in helping people organize information and reflect on creative work \cite{xu2024jamplate}. 
Future work could provide users with templates and schema options to support sense-making in meetings and enhance shared understanding. As suggested in recent work CoExplorer \cite{park2024coexplorer}, the structure of dialogue maps could also be made to adapt to the context of the meeting and change as the discussion evolves to meet various needs of people to externalize verbal communication.   

\subsection{AI-assisted Collaborative Dialogue Mapping for Collaborative Sense-making and Constructive Discussion}
Our study shows that AI-assisted collaborative dialogue mapping can effectively and less confrontationally address misunderstandings. Specifically, AI-generated content serves as a neutral basis for discussions, helping to reduce the personal biases often present in human-generated notes (\S5.1.2).
The AI-generated nodes were also used as scaffolds to organize the subsequent discussion. Unlike previous approaches that positioned AI as an active mediator in group discussion \cite{do2023err, kim2020bot}, our findings indicate that when AI-generated content is designed and presented in a non-intrusive way, it can subtly improve people's metacognitive awareness and guide behavioral changes \cite{tankelevitch2024metacognitive}, for example, motivating users to actively monitor the conversation and actively edit the shared notes.

Besides, creating collaborative notes with AI can facilitate a more balanced and inclusive collaborative environment (\S5.1.4). Participants who used MeetMap actively participated and shared responsibilities more evenly compared to the baseline condition. % as evidenced by the relatively balanced note-taking behavior under MeetMap conditions.
Participants were more willing to edit notes generated by AI than those created by their peers. %This indicates that AI-generated notes %offer a neutral starting point, 
%facilitate easier initiation and development of collaborative efforts. 
This suggests that employing AI-generated content as the basis for collaboration can promote shared ownership and reduce the perceived intrusion of modifying peers' contributions in other collaborative note-taking scenarios \cite{Courtney2022IndividualVC}. Additionally, participants still faced challenges coordinating with each other in MeetMap, including unexpected edits to each other's notes. To address these issues, more explicit mechanisms should be designed to guide collaboration. For example, displaying the edit history on the maps and incorporating features to accept or reject changes made by others could potentially improve coordination \cite{dabbish2012social}. Building on our findings, future research could further investigate the impact of AI-assisted collaborative dialogue mapping on conversation and collaboration dynamics.

\subsection{User Agency and AI Assistance in Sense-making Tasks}
%Using AI to facilitate in-situ dialogue mapping and sense-making of conversations introduces an assistance dilemma \cite{koedinger2007exploring}. Making sense of conversations is a task that demands people's attention and cognitive capacity; it remains a question of what level of AI assistance is desirable and useful. 

Our study suggests that when providing AI assistance to users on cognitively demanding tasks, such as creating dialogue maps to make sense of meetings, the goal should not be to have AI take over the task and generate the final deliverable (in this case, dialogue maps) directly \cite{koedinger2007exploring, kangSynergiMixedInitiativeSystem2023}.  
While AI-generated summaries were well-received, users also valued the control and agency afforded by the Human-Map (\S5.2.1), where they engaged more deeply with the content. The users wanted to make an effort to create the dialogue maps, despite it being more cognitively demanding (\S5.2.3). T
% his desire suggests that when using AI to assist with cognitive tasks such as meeting understanding, while AI can handle mechanical aspects of note-taking, it should not replace the cognitive and creative processes that contribute significantly to user involvement and meeting outcomes. 
Consistent with previous research, our findings indicate that AI is more effective when it complements human cognitive functions rather than replacing them \cite{kangSynergiMixedInitiativeSystem2023, tankelevitch2024metacognitive}. Moreover, our study recommends offering customizable levels of AI assistance to accommodate different user preferences and situational needs (\S5.2.2), e.g., allowing users to choose between more hands-on involvement in organizing content or focusing on refining AI-generated structures.  

%It's important to note that AI-assisted communication aims to improve human communication rather than replace it \cite{zheng2023competent}. 

%- 5.2.1, 5.2.2, 5.2.3

% It is worth noting that 
% may not be well digested by them. Previous studies suggested providing people with partial notes rather than complete notes to help people transfer text knowledge since it can incorporate their own experiences and to help elaborate the new information \cite{katayama1997getting}. This is aligned with the design of the two variants of MeettMap. In both scenarios, users took an active role, organizing the content themselves or re-adjusting AI-generated structures. 

% This highlights the importance of human intervention in capturing the nuances of a discussion. As previous work suggested, users should concentrate more on higher-order thinking behaviors, e.g., focus on algorithmic logic rather than syntactic intricacies in block-based programming \cite{weintrop2019block}, making critical decisions than do laborious information retrieval in human-AI teaming \cite{zheng2023competent}. 

%\subsection{Towards I}
Last, our findings highlight the nuanced dynamics of user engagement with AI-generated content and their varying levels of trust and tolerance towards AI errors (\S5.2.4). Users who actively refined AI-generated content demanded higher accuracy and better results. Moreover, users exhibited a lower tolerance for AI mistakes when they felt a stronger sense of ownership over the collaborative output. An important implication from these observations is that errors in AI-generated content can be tolerable if users are not required to interact extensively with it (e.g, users can refer to the AI-generated dialogue maps in immediate post-meeting reviews). 
However, ensuring AI accuracy becomes paramount when significant participant involvement is needed (e.g., when users actively created the dialogue maps themselves during the meetings).  
In such a scenario, AI must be introduced transparently so that users can easily understand the rationale behind the AI-generated content to fully assess the AI output\cite{wuPromptChainerChainingLarge2022, amershi2019guidelines}.

\subsection{Limitations and Future Work}

\begin{enumerate}
    \item While two proactivity levels of AI were considered in MeetMap, the design does not consider the full spectrum of AI assistance \cite{amershi2019guidelines}. Future work can comprehensively evaluate the role of AI in assisting collaborative note-taking and communication, considering a wider range of AI proactivity levels. 
    \item  We focused on evaluating the user experience in MeetMap through a user study and did not technically evaluate the accuracy of the AI-generated nodes/maps. Future iterations might develop adaptive algorithms, catering to each meeting's dynamic, to produce higher-quality dialogue maps.
    \item The current study of MeetMap focused on new teams where team members did not know each other before. This resembles the experience for many in-class project discussions and ad-hoc workplace meetings \cite{stewart2020beyond}. The results found in this work would apply to new and non-established teams. Additionally, we recruited the evaluation study participants from a mailing list at one university and ran the evaluation study with dyad meetings. In future studies, we hope to recruit a more diverse group of participants and examine the scalability of MeetMap in larger group settings and across different types of organizations, industries, and use cases.   
    %\item  In this study, we did not employ specific metrics or methods to rigorously assess how the use of MeetMap affects the overall comprehension of the meeting content. We plan to analyze the team dynamics in more detail with a longitudinal study. 
    \item We recognize that additional features can add unnecessary burdens. We identified the lack of adaptivity to close unnecessary panels in our system as a limitation. Future systems should allow users to minimize extra cognitive load through more adaptive design.
    \item As shown in Figure \ref{fig:usage}, both note-checking and note-creation behaviors were minimal in the baseline condition, likely due to the high pace and brief nature of the discussion. In such conditions, users may decide not to take notes if they don’t perceive immediate benefits. The two MeetMap variants introduced new interaction mechanisms and AI- scaffolding to encourage collaborative output. In contrast, the baseline setup with Otter.ai transcripts and Google Docs did not provide enough motivation for note-taking. A more suitable baseline might include a built-in collaborative note editor with AI-generated text summaries, which could better validate the usefulness of the design of the collaborative dialogue maps.
    \item The current design uses a low-contrast color scheme, which may not be appropriate for all accessibility requirements. In the future, MeetMap will improve accessibility and visual design by rethinking the color scheme and resizing interactive buttons. These adjustments will help make the system more inclusive and user-friendly.
\end{enumerate}

\section{Conclusion}
Traditional video meeting platforms present discussions linearly, either as transcripts or summaries, while conversation ideas often emerge non-linearly.  We explored LLM-assisted real-time collaborative dialogue map generation, visually representing structured and interconnected ideas. To balance reducing user cognitive load and granting user control over AI-generated content, we introduced two human-AI collaboration approaches: Human-Map and AI-Map. In Human-Map, AI summarizes conversations into nodes, with users linking the nodes to shape a dialogue map. AI-Map allows the AI to create small maps first, which users can subsequently refine. We evaluate these methods through a within-subject study involving ten user pairs. Users preferred  MeetMap over conventional note-taking strategies since it reflected the conversation process and aligned with how humans organize information. Users liked the ease of use for AI-Map due to the low effort demands and appreciated the
hands-on opportunity in Human-Map for sense-making. This study provides insights on enhancing human-AI collaboration in note-taking and sense-making during meetings and provides design implications for involving AI in synchronous human communication activities. 

\section*{Acknowledgments}

This material is based upon work supported by the National Science Foundation under Grant Number IIS-2302564.

\bibliographystyle{ACM-Reference-Format}
\bibliography{references}

\appendix
\section{Appendix: survey questions and statistical results}
\label{cognitive-load-survey}
\begin{figure}[htbp]
    \centering     
    \includegraphics[width=1.0\textwidth]{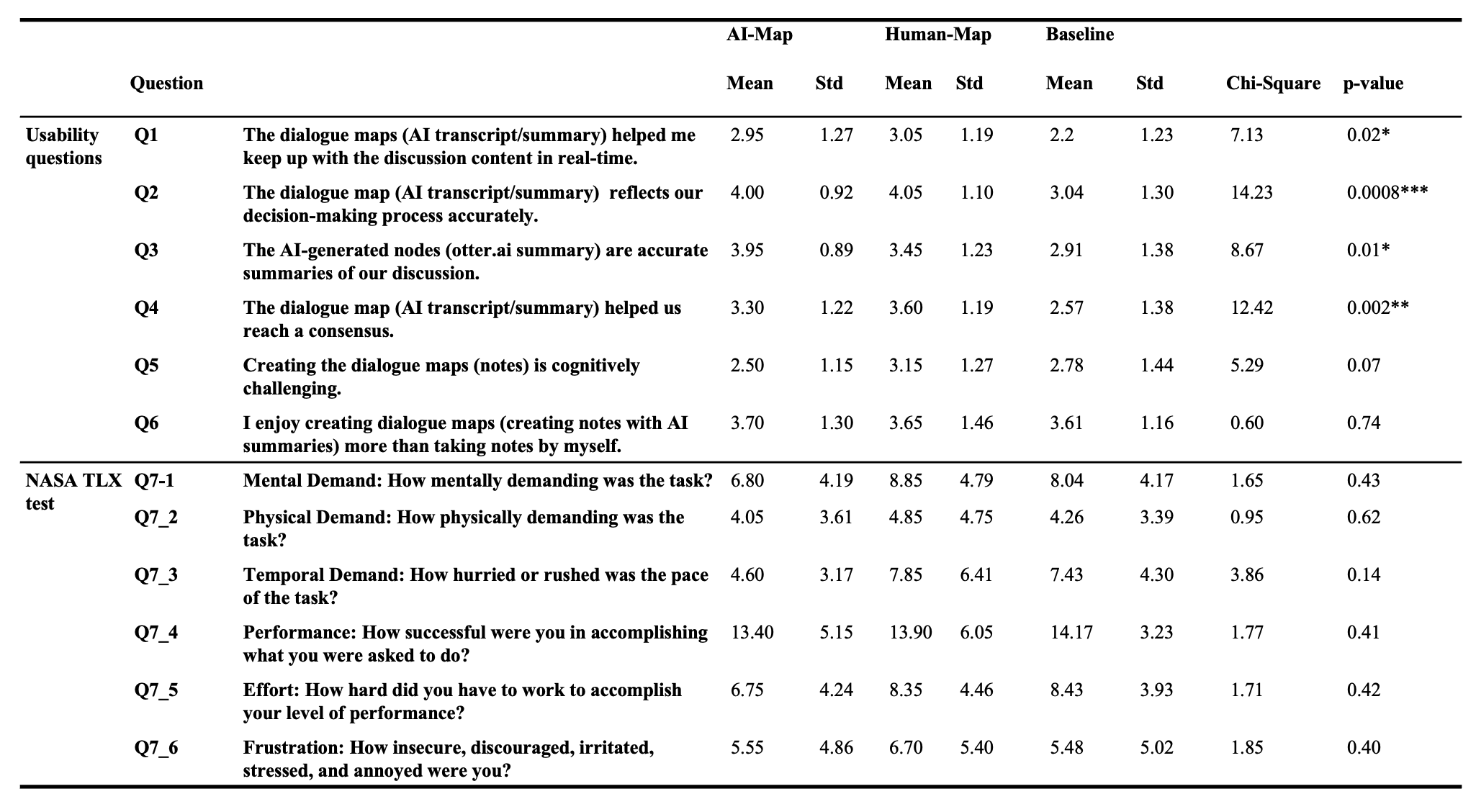}
    \caption{\textbf{Descriptive and statistic analysis of the survey questions} ($*: p<0.05$; $**: p<0.01$; $***: p<0.001$)}
    \label{fig:survey}
\end{figure}
\begin{figure}[htbp]
    \centering     
    \includegraphics[width=\textwidth]{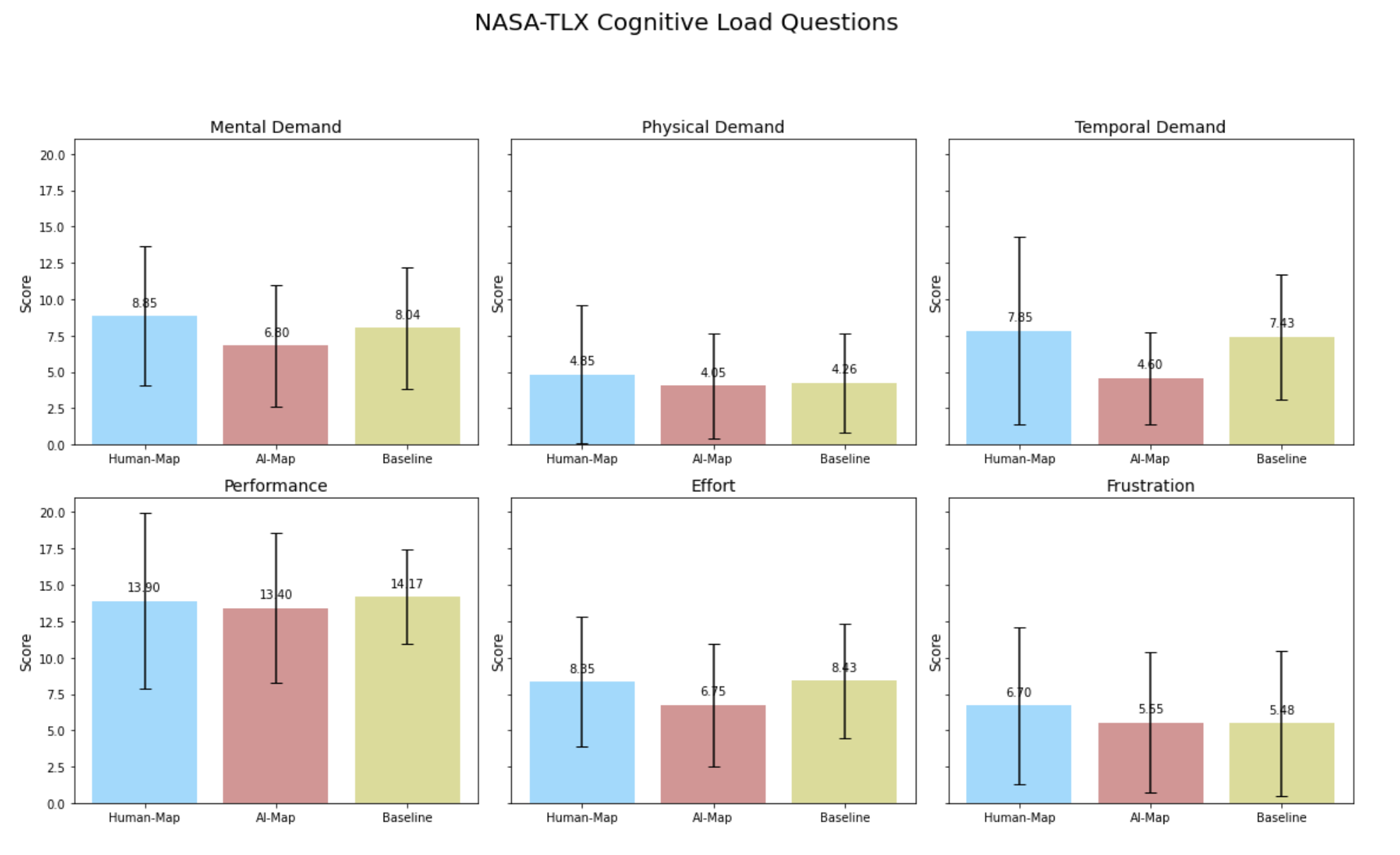}
        \caption{Cognitive load across the conditions ($*: p<0.05$; $**: p<0.01$; $***: p<0.001$)}
    \label{fig:cognitiveload}
\end{figure}
\label{survey}
\section{Appendix: Semi-structured Interview Questions}\label{InterviewQuestion}

\subsection*{Questions after Each Condition (\HumanMap, \AIMap, \Baseline)}

\begin{enumerate}

    \item Usefulness of AI aids
    \begin{enumerate}
        \item Was the AI aids helpful for you? How? Why or why not?
        \item Were there moments when the AI aids helped you understand the discussion better? Can you tell me more about it?
        \item How important is it that these AI aids are shown to you in real-time?
    \end{enumerate}

    \item Cognitive Load During Discussion with Real-Time Aids
    \begin{enumerate}
        \item Did you read the temporary AI aids during the discussion? If yes, how did that influence your discussion?
        \item Did you refer back to the AI aids during the discussion? If yes, how did that influence your subsequent discussion?
    \end{enumerate}

    \item Thoughts on the Break
    \begin{enumerate}
        \item How did you distribute your time between creating/reading the AI aids during the discussion and the break?
        \item Would you prefer to interact with the AI aids during the discussion or during the break? Why?
    \end{enumerate}

    \item Use of AI aids in the Reflection Period
    \begin{enumerate}
        \item Did you use the AI aids when answering the survey? How did you use them?
    \end{enumerate}

    \item Quality of AI aids
    \begin{enumerate}
        \item What are your thoughts on the quality of the resulting AI aids?
        \item What are your thoughts on the quality of the AI-generated nodes (if applicable)?
    \end{enumerate}

    \item Challenges
    \begin{enumerate}
        \item Can you share some challenges you encountered during the discussion? 
    \end{enumerate}

\end{enumerate}

\subsection*{ Comparison Between the Two Dialogue Mapping Conditions (\HumanMap and \AIMap)}

\begin{enumerate}

    \item Comparison of Cognitive Demands
    \begin{enumerate}
        \item Can you share the cognitive demands on you for both dialogue mapping methods (\HumanMap and \AIMap)?
    \end{enumerate}

    \item Comparison of Quality of Dialogue Maps
    \begin{enumerate}
        \item Can you compare the quality of the resulting dialogue maps in both methods?
    \end{enumerate}

    \item Trustworthiness of the Methods
    \begin{enumerate}
        \item Can you share your thoughts on the trustworthiness of the two dialogue mapping methods?
    \end{enumerate}

    \item Preference for Future Meetings
    \begin{enumerate}
        \item Which dialogue mapping method (\HumanMap or \AIMap) would you prefer for a meeting in the future? Why?
    \end{enumerate}

\end{enumerate}

\subsection*{Comparison Between All Three Conditions (\HumanMap, \AIMap, and \Baseline)}

\begin{enumerate}

    \item Cognitive Demands Across Conditions
    \begin{enumerate}
        \item When you were creating the dialogue maps (\HumanMap, \AIMap) versus taking notes yourself (\Baseline), which one was more cognitively demanding for you?
    \end{enumerate}

    \item Helpfulness for Visualizing Discussion Progress and Decision-Making
    \begin{enumerate}
        \item Comparing the resulting dialogue maps (\HumanMap, \AIMap) with the AI summaries and transcripts (\Baseline), which was more helpful for you to visualize the discussion progress and make decisions?
    \end{enumerate}

    \item Quality of Summaries and Dialogue Maps
    \begin{enumerate}
        \item Can you comment on the quality of the summaries (\Baseline) and the quality of the dialogue maps (\HumanMap, \AIMap)?
    \end{enumerate}

    \item Preference for Future Meetings
    \begin{enumerate}
        \item Which method (\HumanMap, \AIMap, or \Baseline) would you prefer for a meeting in the future? Why?
    \end{enumerate}

\end{enumerate}

\section{Appendix: Tasks for User Study}
\label{task}

Each task was presented to participants in dyads, with each participant viewing content tailored to their assigned persona. The specific tasks, agendas, and personas for each user are provided below.

\subsection{Task 1 - Enhancing Mental Health Services on Campus}

\textbf{Task Description:}  
Student representatives are tasked to discuss and formulate strategies to improve mental health and well-being support on campus.

\textbf{Agenda 1:}  
One strategy is to require first-year students to pay a 15-minute visit to the University Counseling \& Psychological Services (CAPS) office as a prerequisite for class registration.
Please discuss whether you agree with this strategy and reach a consensus on what is the best way to go:

- \textbf{Persona 1: You support this strategy}  
    - There is an increase in reported student anxiety, stress, and other mental health issues. However, many students don’t seek help from CAPS until the problems become serious.
    - A required visit to CAPS can help students get familiar with the services they provide.

- \textbf{Persona 2: You do not support the requirement}  
    - If this becomes a requirement, it will be a big load for the CAPS office, and they will need to increase staff.
    - Students who do not need support may find this to be a waste of time.

\textbf{Agenda 2:}  
Given the diverse student population, what mental health services would you consider essential? Discuss their significance and decide which one should be prioritized. Please propose your solutions. If you run out of ideas, here are some suggestions.

\begin{itemize}
    \item \textbf{24/7 Helpline with student volunteers:}
    \begin{itemize}
        \item + Provides instant support during crises.
        \item + Can be accessed anonymously.
        \item - Needs training for student volunteers to provide counseling services.
        \item - Student volunteers may not be as effective as professional counselors.
    \end{itemize}
\end{itemize}

\subsection{Task 2 - Installing Smart Devices for University Buildings}

\textbf{Task Description:}  
Student representatives are tasked to discuss installing smart devices for a new university building (e.g., facial recognition devices for building entrances, smart lighting systems, etc). Your goal is to share your perspectives on what smart devices to install and reach a consensus.

\textbf{Agenda 1:}  
To support easy entrance into buildings, what smart techniques may you consider? Try to generate as many ideas as possible. Please propose your solutions. If you run out of ideas, here are some suggestions.

- \textbf{Persona 1: Budget-focused perspective}
    \begin{itemize}
        \item Scanner at the entrance that scans QR code on the phone for identification
        \begin{itemize}
            \item + Students will most likely have phones with them
            \item + Cheap and easy (no energy cost)
            \item - Insecure, people can share QR codes with anybody
            \item - Phones could be out of battery
        \end{itemize}
    \end{itemize}

- \textbf{Persona 2: Technology proponent perspective}
    \begin{itemize}
        \item Facial recognition system using computer vision
        \begin{itemize}
            \item + Students don’t need to bring anything with them
            \item + Quick and easy
            \item - Expensive
            \item - AI errors
        \end{itemize}
    \end{itemize}

\textbf{Agenda 2:}  
To support occupancy sensing so that students can know which classrooms and spaces are empty from a dashboard, what smart devices should you consider? Try to generate as many ideas as possible. Please propose your solutions. If you run out of ideas, here are some suggestions.

\begin{itemize}
    \item \textbf{Persona 1: Environmental perspective}
        \begin{itemize}
            \item Occupancy sensors that detect motions in a room
            \begin{itemize}
                \item + Cheap
                \item + Less privacy concerns 
                \item - Less information, e.g., don’t know how many people
                \item - May send errors when there are people but no movements
            \end{itemize}
        \end{itemize}
        
    \item \textbf{Persona 2: Technology proponent perspective}
        \begin{itemize}
            \item Facial recognition system using computer vision
            \begin{itemize}
                \item + Provides more information, e.g., how many people are in a space 
                \item + Robust to many contexts, classrooms, cafes, etc.
                \item - Privacy concerns 
                \item - Expensive
            \end{itemize}
        \end{itemize}
\end{itemize}

\subsection{Task 3 - Reevaluating Attendance Checking in University Classes}

\textbf{Task Description:}  
Student and faculty representatives are tasked to discuss the merits and concerns about attendance checking in classes at the University. You will each represent a student representative and a faculty representative. Your goal is to reach a consensus on the best way to check attendance so that every stakeholder is happy with it.

\textbf{Agenda 1:}  
Please reach a consensus on whether in-person attendance is necessary or not.

- \textbf{Persona 1: Instructor Representative}
    \begin{itemize}
        \item In-person attendance will increase student engagement.
        \item Enhanced peer interactions and community-building.
        \item Higher learning benefits since students will be more active (e.g., more Q\&A) during in-person classes.
    \end{itemize}

- \textbf{Persona 2: Student Representative}
    \begin{itemize}
        \item Difficulties of transportation between north and central campuses.
        \item Daytime jobs or other responsibilities that conflict with class timings.
        \item Enrollment in consecutive or overlapping classes.
    \end{itemize}

\textbf{Agenda 2:}  
What are alternative ways to check participation and attendance? Please reach a consensus on one desirable way to check for student participation and attendance. Please propose your solutions. If you run out of ideas, here are some suggestions.

\begin{itemize}
    \item \textbf{Both personas may consider:}
    \begin{itemize}
        \item Give students several missed opportunities, e.g., they can miss three in-person classes. Otherwise, it is required.
        \item Students can answer a survey within a time window (e.g., within 48 hours) so that they can watch the lecture recording after the lecture time.
    \end{itemize}
\end{itemize}

\section{Appendix: Demographic information of participants}
\begin{table}[H]
\centering

\begin{tabular}{>{\centering\arraybackslash}m{0.05\textwidth}|>{\centering\arraybackslash}m{0.15\textwidth}|>{\centering\arraybackslash}m{0.15\textwidth}|>{\centering\arraybackslash}m{0.20\textwidth}|>{\centering\arraybackslash}m{0.20\textwidth}|>{\centering\arraybackslash}m{0.15\textwidth}}
\hline
\textbf{ID} & \textbf{Gender} & \textbf{Age} & \textbf{Ethnicity} & \textbf{School Year} & \textbf{Online Meetings/Week} \\
\hline
P1 & Male & 21 - 30 & Asian & Master's student & 5-10 \\
P2 & Female & 21 - 30 & White & Master's student & $>$10 \\
P3 & Female & 21 - 30 & Asian & Master's student & 5-10 \\
P4 & Male & 10 - 20 & White & Senior & $>$10 \\
P5 & Female & 30 - 40 & White & Master's student & 1-4 \\
P6 & Female & 21 - 30 & Asian & Master's student & 1-4 \\
P7 & Male & 21 - 30 & Black & Ph.D. student & 5-10 \\
P8 & Female & 21 - 30 & Asian & Junior & 1-4 \\
P9 & Male & 10 - 20 & Asian & Sophomore & 1-4 \\
P10 & Female & 21 - 30 & White & Master's student & 1-4 \\
P11 & Male & 21 - 30 & Black & Master's student & $>$10 \\
P12 & Male & 21 - 30 & White & Master's student & 1-4 \\
P13 & Female & 21 - 30 & Black American & Master's student & 1-4 \\
P14 & Male & 21 - 30 & Black & Master's student & 1-4 \\
P15 & Female & 21 - 30 & Black American & Master's student & 1-4 \\
P16 & Male & 21 - 30 & Black American & Ph.D. student & $>$10 \\
P17 & Male & 21 - 30 & Middle Eastern & Master's student & 5-10 \\
P18 & Female & 21 - 30 & African American & Senior & 1-4 \\
P19 & Female & 21 - 30 & White & Master's student & 1-4 \\
P20 & Female & 31 - 40 & White & Ph.D. student & 5-10 \\
\hline
\end{tabular}
\caption{Demographic Information of Participants}
\label{tab:demographics}
\end{table}

\label{demographic}
% \section{Appendix: interview protocol}

% \label{interview}

\section{Appendix: LLM Prompts} 
\label{appendix:prompts}
\subsection{Topic Segmentation}
%Your role is a conversational topic analyzer. Given a meeting dialogue, identify and categorize the topics. You'll get the most recent exchange and an existing topic list. The topic list will be vacant at first. For each new exchange:\\
%1. Determine the nature of each exchange in the dialogue: Is it a 'Continuation' (Tagged as \$C-Continuation) of the previous topic, Or, is it a 'New Topic' (Tagged as \$N-New)? \\
%2. If it is a continutation, modify the last topic to accomadate for the new turn. The updated topic should not be over 6 words. \\
%3. Should there be related topics in the topic list highly closely connected with this newly added turn, display the related topic clusters (show maximum 3).
%Generate an output that includes the identified topic (under 6 words), the tag (either \$C-Continuation or \$N-New), the related topic cluster (if any), and the updated or newly added topic tag [\$Updated] or [\$Add] and the updated topic/newly added topic. Your output should strictly be in the following JSON format with no additional text or explanations outside of it:\\
%\{
    %"Identified topic": "",
    %"Continuation/New Topic Tag": "",
    %"Related topic clusters": [1, 2],
    %"Update topic or not": [\$Updated]
    %"Updated/Newly added topic": {"key": 2, "topic": "", "turn": [1, 3]}
%\}

Your role is a conversational topic analyzer. Given a meeting dialogue, identify and categorize the topics. You'll get the most recent exchange and an existing topic list. The topic list will be vacant at first. For each new exchange:
\begin{itemize}
    \item 1. Determine the nature of each exchange in the dialogue: Is it a 'Continuation' (Tagged as \$C-Continuation) of the previous topic, Or is it a 'New Topic' (Tagged as \$N-New)? 
    \item 2. If it is a continuation, modify the last topic to accommodate for the new turn. The updated topic should not be over six words), and the tag should be either \$C-Continuation or \$N-New. Your output should strictly be in the following JSON format with no additional text or explanations outside of it:
\end{itemize}

\{
    "Identified topic": "",
    "Continuation/New Topic Tag": "",
\}

\subsection{Dialogue Tagging and Summarization}
You are a facilitator trying to generate dialogue mapping for the conversation. The user will input one turn, and you will analyze the data following the steps below:
\begin{itemize}
    \item 1. Assign dialogue mapping tags to dialogue chunks according to the dialogue mapping schema: [\$Question], [\$Position],[\$Pro], [\$Con]. It captures Questions (or Issues): What are we trying to solve? Ideas (or Responses): Potential answers or solutions. Arguments: Pros and cons for each idea.
    \item 2. Assign a dialogue mapping tag like [\$Question] and find the corresponding sentences shown in one turn, then summarize the content under six words. Please note that each turn might have one or more tags connected to different sentences, or there is no tag if the turn is about back-channeling, greetings, or talking about some non-argumentive stuff. Please ignore the totally off-topic conversation about the decision-making tasks people are working on. Please ensure that the summarized content cannot be divided into smaller pieces under each dialogue tag. For example, if the summarized content is "discuss two devices, automatic lighting, the cardless entry, "it should not be divided into one question tag with two idea tags, each indicating one device.
\end{itemize}
The final output should only be a JSON string, and please do not provide any other text or explanation outside of the JSON format:\\
\{
"dialogueTagArray":[
\{"Tag": "[\$Question1]", "Summary": "Invitation to discuss products," "Quotes": "Does anyone want to talk about their products?"\},
],
\}

\subsection{Link Identification}
You are a facilitator trying to generate dialogue mapping for the conversation. The user will input a list of nodes with dialogue tags.  The inputted nodes were assigned with a key; please do not change it. You will analyze the data following the steps below:
\begin{itemize}
    \item 1. Please identify relationships between nodes in the nodes list, such as [\$Positions] answering [\$Question], [\$Pros] supporting [\$Position]. And generate a comprehensive hierarchy node structure for the conversation chunk. 
    \item 2. The link should be in a single direction; this means one node should only be one time as a "from" node, but others can link one node many times.
\end{itemize}
The final output should only include a JSON data structure of the links like the following, and please do not provide any other text or explanation outside of the JSON format:\\
\{
"linkDataArray": [
\{ "from": 2, "to": 1, "text": "Support"\}
]
\}

\end{document}